\documentclass[12pt,preprint]{aastex}

\def\etal{et~al.}
\def\eg{e.g.}
\def\deg{^\circ}
\newcommand{\hii}{{H~{$\scriptstyle {\rm II}$}}}
\newcommand{\hi}{{H~{$\scriptstyle {\rm I}$}}}

\begin{document}
\title{G28.17+0.05: An Unusual Giant \hi\ Cloud In The Inner Galaxy} 

\author{Anthony H. Minter, Felix J. Lockman, Glen I. Langston}
\affil{National Radio Astronomy Observatory
\footnote{The National Radio Astronomy Observatory is a facility of the 
National Science Foundation operated under cooperative agreement with
Associated Universities, Inc.}, P.O. Box 2, Green Bank, WV, 24944
; tminter@nrao.edu, jlockman@nrao.edu,glangsto@nrao.edu}
\authoraddr{P.O. Box 2, Green Bank, WV, 24944}
\and
\author{Jennifer A. Lockman \footnote{Current address
University of Minnesota, Department of Astronomy, Minneapolis, MN 55455}}
\affil{College of Charleston,Department of Physics, Charleston, SC, 29424;
jlockman@astro.umn.edu}
\authoraddr{Department of Physics, Charleston, SC, 29424}

\begin{abstract}
New 21 cm \hi\ observations have revealed a giant \hi\ cloud in the 
 Galactic plane that has unusual properties.  It is quite well defined, 
about 150 pc in diameter at a distance of 5 kpc, and  contains 
as much as 10$^5 M_{\sun}$ of atomic hydrogen.  The outer parts of 
the cloud appear in \hi\ emission above the \hi\ background, 
while the central regions show \hi\ self-absorption.  
Models which reproduce the observations have a core 
with a temperature $\lesssim 40$ K and
an  outer envelope as much as an order of magnitude  hotter.
The cold core is elongated along the Galactic plane, whereas the
overall outline of the cloud is approximately spherical.
 The warm and cold parts of the \hi\ cloud have a similar, and 
relatively large, line width $\sim 7$ km~s$^{-1}$.  
  The cloud core is a source of weak, anomalously-excited 1720 MHz OH emission,
also with a relatively large line width, 
which delineates the region of \hi\ self-absorption but is slightly
blue-shifted in velocity. The
intensity of the 1720 MHz OH  emission is correlated with N$_H$ 
derived from  models of the cold core.    
 There is $^{12}$CO emission associated with the
cloud core.  Most of the cloud mass is in molecules, and
the total mass is $> 2 \times 10^5 M_{\sun}$.  
 In the cold core the \hi\ mass fraction may be $\sim 10\%$. 
The cloud has only a few sites of current star formation.
There may be  $\sim 100$ more objects like this in the inner 
 Galaxy;  every line of sight through the Galactic plane within
$50\deg$ of the Galactic center probably intersects at least one.
 We suggest that G28.17+0.05 is a cloud being observed as it 
enters a spiral arm and that it is in the transition from the atomic to 
the molecular state.
\end{abstract}

\section{Introduction}
Studies  of Galactic \hi\ have generally 
been made either with single antennas, which
provide information on the \hi\ in emission at fairly low angular
resolution, or with interferometers, which  
until recently, were used mainly to  observe Galactic \hi\ at
high angular resolution 
in absorption against background continuum sources (\eg\ 
\citet{heiles67,clark}; see reviews in \citet{hk,burton,dl90}). 
When Galactic \hi\  was able to be observed at
 an angular resolution of several minutes of arc, 
 it was discovered that much of the structure in the 
\hi\ emission profiles at low
Galactic latitude is caused by self-absorption from cool atomic clouds
being viewed against a brighter \hi\ background 
(\citet{baker,liszt81,bl84,liszt84}).
Low-latitude \hi\ 
emission profiles are a mix of emission and self-absorption.  
The angular resolution needed  to discern 
the cold clouds seen in self-absorption depends on their 
size and distance, and perhaps it is not surprising that 
existing catalogs of absorbing
clouds are not representative, but are 
dominated by objects with  a size of 
a few resolution-elements (\eg\ \citet{bl84}).
 New aperture synthesis and  single-dish surveys 
are finally providing high-resolution data on portions of the Galaxy
\citep{bk90,higgs,gibson},
but the inner $30\deg$ of the Galactic plane, which contains
most of the star-forming regions and molecular clouds in the Galaxy, still 
has not been observed in \hi\ at  better than $35\arcmin$ resolution, 
and even at that resolution, sampling is not complete.  

This lack of data has seriously compromised our understanding of
basic processess.  The neutral interstellar medium contains
a spectrum of objects, from from diffuse  \hi\ 
clouds of a solar mass or less, to the giant molecular clouds and cloud 
complexes which can exceed $10^6\  M_{\sun}$ and are the sites of the 
formation of the most massive stars (see \citet{turner97} for a recent
review).  The relationship between these objects is unclear, but 
material is observed to move between the various phases.   
Many molecular clouds are surrounded by \hi\ halos, and,  conversely, 
molecules are
observed in the densest parts of predominently atomic clouds.  	It is 
thought that atomic \hi\ clouds can undergo a phase
transition at a spiral density wave-induced shock to form giant molecular
clouds, which are then disrupted after some time by the products of their
own star formation (\eg\
\citet{allen86,elmegreen95}).
This paper describes the discovery of an unusual cloud which may 
be in a transitional state between  atomic and molecular.

\section{A Hole in the Galactic Continuum Emission}

Our attention was first drawn to 
the area near longitude $28\deg$ by the information in Figure 1, 
which shows  spatially-filtered 8.35 GHz continuum emission
 for a portion of the Galactic plane. 
These observations were made at $11.2 \arcmin$ resolution in a 
search for transient radio sources as part of the GPA survey \citep{gpa}.
The data have been passed through a spatial filter in Galactic latitude to 
suppress emission on scales $> 1 \deg$.  This 
reduces zero-level drifts in the data, but also results in the depressions
that run through Figure 1 above and below the Galactic plane.

There is a large, cone-shaped minimum  in the Galactic continuum centered
near $27\fdg8$ longitude.   The hole has no detectable 8.35 GHz continuum 
emission down to the $\sim0.1$~Jy flux limit of the (spatially filtered) 
GPA survey and it is bounded on all sides by relatively bright 
 emission.  This hole in the continuum is also 
apparent in the 4.875 GHz survey of 
\citet{altenhoff}, which was made 
 using the 100 meter telescope of the Max Planck Institute for Radio Astronomy.
While there is no known object at the location of the hole, 
the bright emission which caps the hole has been 
identified as a supernova remnant (SNR) G27.8+0.6.  There is also another
SNR adjacent to the hole at lower longitudes, G27.4+0.0
\citep{green}, and a number of 
\hii\ regions on either side of the hole 
\citep{jay,diffuse}.
While we now believe that this hole in the continuum has 
little significance, it caused us to make \hi\ measurements which did
reveal something quite unusual.

\section{\hi\ Observations}

To determine if the radio continuum feature had any counterpart 
 in atomic gas, the region $26\fdg5 \le l \le 28\fdg65$, 
$|b| \le 1\deg$ was observed in the  21 cm line of \hi\ using the 
NRAO 140 Foot Telescope which has a 
half-power beam-width (HPBW) of $21\arcmin$ at the
frequency of these observations. Spectra with a velocity resolution
of $1.0~{\rm km~s^{-1}}$ covering 
${\rm -300~km~s^{-1} < V_{LSR} < +225~km~s^{-1}}$ 
were taken every $8\arcmin$  in $l$ and $b$.
These are the highest angular resolution 21cm \hi\ 
observations ever made of this
part of the Galaxy, except along a narrow strip at $b=0\deg$.
 The \hi\ spectra were corrected for stray radiation
using the all-sky deconvolution method of \citet{kalberla} as applied
to the 140 Foot Telescope by \citet{murphy}.  This procedure 
also places the \hi\ intensities on the absolute brightness temperature
scale defined by \citet{kalberla82}.  
As a  supplement to the 140 Foot
data, H.S. Liszt has kindly provided us with a set of 
unpublished \hi\ spectra
taken by G. Westerhout using the 100 meter radio telescope of the 
Max Planck Institute for Radio Astronomy in Bonn, Germany (hereafter
the ``100 meter'' data).  These spectra were taken at 
 $b=0\deg$ every $5\arcmin$ in Galactic longitude with 
 a $9\arcmin$ angular resolution
and $2.0~{\rm km~s^{-1}}$ velocity resolution. 
The 100 meter data have not been
corrected for stray radiation and their absolute calibration is consequently 
uncertain, so they were used mainly  to  check  model predictions 
at $b=0\deg$ (see \S 4.2).

 Figure 2 shows the \hi\ brightness temperature from the 140 Foot 
observations as a function of 
Galactic longitude and latitude for a set of velocities 
 from ${\rm V_{LSR}=72.6~km~s^{-1}}$ to ${\rm 88.0~km~s^{-1}}$.
There is a shell surrounding a hole in the \hi\ centered  at 
$(\ell,b) = 28\deg, 0\deg$
approximately coincident with the hole in the radio continuum. 
Initially, we believed that the feature might be a cavity with a 
shell, perhaps the result of a supernova.   
Spectra towards the center of the \hi\ hole, however, 
suggest not an absence of gas, but rather \hi\ self-absorption.
Figure 3 shows the 140 Foot \hi\ 
spectrum at $28\fdg00+0\fdg00$.  The arrow marks the velocity
of the \hi\ hole, which appears in the spectrum as 
a minimum with a FWHM of about 7 km~s$^{-1}$  at a velocity of 79 km~s$^{-1}$.
  A feature like this with such steep edges would be 
 difficult to produce by a void in space in the inner Galaxy:
random motions of clouds and the relatively small change in
projected velocity with distance over most of the Galaxy creates a
confused background which generally 
obscures true density minima (see the discussion
in \cite{baker}). Perhaps most important, 
 Figure 3 shows that there is  molecular gas
with nearly the same velocity and line width as the \hi\ minimum,
an unlikely occurrence if the minimum is a void rather than self-absorption.

The object we have found is most likely an \hi\ shell surrounding a  
concentration of \hi\ so cold and
opaque that it absorbs \hi\ emission from warmer gas behind it. 
Toward the cloud center the \hi\ brightness temperature, T$_b$, 
falls as low as 50 K, while the \hi\ brightness temperature 
at that same velocity less than a degree away in longitude 
in either direction is $\sim 70$ K.  At its brightest, 
the \hi\ shell is enhanced above the background \hi\ by about the 
same amount as the self-absorbed core is depressed below it.
Figure 4 shows a cut through the data at $b = 0\deg$ 
illustrating the bright rim and depressed emission in the central region.  
The crosses are the \hi\ data and the solid line the
results of a model (\S 4.2).

\hi\ self-absorption is seen throughout the inner Galaxy and in dark
clouds all over the sky 
\citep[and references therein]{knapp,mccutch,baker,liszt81,bl84,gibson,dl90}.
The geometry of self-absorption
requires the absorbing cloud to be silhouetted against a 
background of brighter emission at the same velocity.  
The velocity of the deepest absorption is 
79 km~s$^{-1}$, which corresponds to a kinematic distance of
either 5 or 10 kpc for a flat rotation curve of 220 km~s$^{-1}$
and a Sun-center distance of 8.5 kpc.
A spectral feature arising from 
 self-absorption can be filled in by emission from 
foreground gas, so detection of a strong feature like this 
argues that the cloud lies at the near kinematic distance
\citep{baker,bania00}.  
Throughout this paper we will assume that the
 cloud is at the near kinematic distance of 5 kpc from the Sun,   
which places it 4.8 kpc from the Galactic center.

The velocity of the cloud varies slightly with longitude
with a magnitude consistent with 
the changing projection of Galactic rotation, and 
the central velocity 
of the absorption feature is well approximated by 
${\rm 79\ sin(\ell)/sin(28\fdg2)\  km~s^{-1}}$.
At the highest longitudes there
is some evidence for the presence of 
another \hi\ cloud at somewhat higher velocity
in addition to the main feature.  This is at the edge of our 
map and we will not consider it further.  

\section{Properties of the \hi\ Cloud}
\subsection{General Considerations}

 The self-absorbed lines measured by the 140 Foot 
have a full-width to half-maximum, $\Delta v$, 
that ranges from 6 to 10 ~km~s$^{-1}$
with an uncertainty of several km~s$^{-1}$ arising from uncertainties
in the estimate of the HI emission level in the absence of absorption. 
At the cloud center $\Delta v \sim {\rm 7~km~s^{-1}}$.  The 100 meter
data give similar values.   For 21 cm lines of 
this velocity width, the peak optical depth $\tau = 0.74\ 
{\rm N_{19}\ T_{ex}^{-1}} $ where ${\rm N_{19}}$ is the \hi\ column
density in units of $10^{19}$ cm$^{-2}$ and ${\rm T_{ex}}$ is the
excitation temperature in Kelvins.
The cloud must have an 
optical depth of at least unity in order to be opaque to background
emission, so the column density of \hi\ through the cloud must be 
 ${\rm N_{19} > T_{ex}}$, 
and the excitation temperature is limited by the 
brightness temperature of \hi\ at the location of maximum 
absorption to  ${\rm T_{ex} \lesssim 50}$ K.  The self-absorption profiles
appear to be nearly Gaussian, so the optical depth is not large enough
to cause appreciable saturation in the line.

The \hi\ emission from the outer parts of the cloud, which we will call 
the shell, is as much as 25 K above the background 
emission.  It is especially difficult to determine a line-width for emission  
from the shell because of blending with other \hi, but we estimate 
 that $\Delta v \sim $ 8 to 10  km~s$^{-1}$, similar to the 
6 to 10 km~s$^{-1}$ found for the self-absorbed lines from the core.
To produce the observed emission the 
 shell must have a column density of a few $10^{20}$ cm$^{-2}$.
 \citet{sanbonmatsu} have detected a $\tau \approx 0.1$ \hi\ absorption 
line at the velocity of the cloud against the radio 
SNR G27.4+0.0, which lies  projected on a portion of  the shell 
at a distance of 6-7.5 kpc.   This observation  supports the 
choice of the near kinematic distance for the \hi\ cloud, and, given
the brightness temperature of the shell emission, implies that 
 ${\rm T_{ex} \approx 250}$ K in the shell.

As can be seen in Figure 2, 
the extent of the cloud is approximately $2\deg$ in longitude and
$1\fdg5$ in latitude, giving a size of about $175 \times 130$ pc at
a distance of 5 kpc.  
For an average \hi\ column density of $5 \times 10^{20}\  {\rm cm^{-2}}$,
the total implied \hi\ mass is $7 \times 10^4\  M_{\sun}$.
These general considerations 
suggest that we have discovered a large, exceptionally massive, \hi\ cloud,
with a temperature change of perhaps an order of magnitude between
the core and the shell.  Analysis of \hi\ self-absorption features can
involve some subtleties (e.g. \cite{levinson}).  Antenna beam
convolution is a specially important factor for this cloud, 
so to derive more reliable
estimates of the cloud properties we have calculated a model
which attempts to reproduce the observations taking into account all
relevant effects.

\subsection{A Detailed Model}

To make a more precise estimate of the properties of the \hi\ cloud,
 we have approximated it using an extremely simple model: 
 a spheroid with semi-major axes $\rm{ r_x,\ r_y,\ and\ r_z}$,
in directions parallel to the Galactic plane, along the line of
sight, and perpendicular to the Galactic plane, respectively.  
 Models with a linear temperature gradient 
 did not successfully fit the data --  the steep change in T$_b$
between core and shell requires a nearly discontinuous
change in  temperature.
A simple two-component model was thus adopted with
  an outer shell and inner core having 
 constant, but differing, excitation temperature. 
The density was allowed to be continuously and linearly variable from
center to edge.  The line width was assumed to  
be 7 km~s$^{-1}$ (FWHM) throughout,
as the observations suggest that this value is an appropriate 
mean for both 
the core and shell.   The core has its own 
semi-major axes, but its center is assumed to lie at 
the center of the overall cloud.
The core is thus merely the region of low temperature in a cloud
of otherwise smoothly varying properties.  This simple model is 
sufficient to account for most features of the \hi\ data.

Our calculations evaluated the equation of radiative transfer
for an \hi\ cloud  observed against a background
of unrelated \hi\ emission.  The background was determined from the 
average of data at longitudes 26\fdg78 and 26\fdg91, which are
just beyond the edge of the cloud, and
 was modeled by a Gaussian function  with a dispersion of 60
pc in z, a peak brightness temperature 
in \hi\ of 71 K, and an offset with respect to 
the Galactic plane of $-10$ pc.  

Most models were calculated using
 $5 \times 5$ pc cells in x and z (corresponding to  3.4 arcmin 
in longitude and latitude) and 1 pc cells along
the line of sight.  Within each cell the temperature and
density were  constant.  The emergent spectrum was calculated 
as a function of velocity offset from the line center, 
 at a velocity interval of 1 km~s$^{-1}$, with the assumption that
the model cloud has no rotation or other systematic motion besides
Galactic rotation.    The emergent brightness 
temperatures from the model were  convolved with a two-dimensional Gaussian 
to match the angular resolution of the observations.
The models were created and evaluated using the 
AIPS++ software package.

Properties of representative  models that fit the data 
 are shown in Table 1.  The cloud outer
radius in the x direction (parallel to the Galactic plane) varies
slightly, but for 
all models shown here the semi-major axes r$_y =$ r$_z = 75$ pc.  
The semi-major axes of the cold core are given in columns 3-5.  The average 
\hi\ volume density varies linearly between $n_0$, its value 
at the center of the cloud, and $n_r$, its value at the cloud edge.
Column 10 gives the displacement from the Galactic 
plane of the cloud center; the total \hi\ mass in column 11 does not include
He or other elements.

Figure 5 shows the  \hi\ brightness temperature averaged over the  
velocity range of the cloud along strips 
perpendicular to the Galactic plane at three longitudes.
 Two of the strips cut through the cloud shell and 
core while the third intersects only the shell.  
The solid line shows the calculations for Model I,
 the model which best fits the data.  Figure 4 shows the data 
along a cut 
parallel to the Galactic plane, also with the results from Model I.
The models place the cloud
center at $\ell = 28\fdg17$  and just slightly 
(2 to 5 arcmin) above the Galactic plane.  
We adopt a cloud center of $28\fdg17+0\fdg05$.  
The maximum column density through the center of the models is between 
0.5 and  $1.2 \times 10^{21}$  cm$^{-2}$, and the core contains 
between 20\% and 40\% of the total \hi\ mass.
All models discussed here
gave an optical depth within a factor of two of that observed towards
the SNR 27.4+0.0 \citep{sanbonmatsu}.  Although the 140 Foot
data do not completely cover the cloud at the higher longitudes, 
the 100 meter data at $b=0\deg$  are in 
reasonable agreement with the models and indicate that not much of the
cloud has been missed.

Models in which  the cold core  extends all the
way to the cloud surface at low z, as in all but Model IV, 
give the best fit to the data, and have the highest core 
temperature and largest total \hi\ mass.  Core temperatures
below 40 K can only be reconciled with the observations if there
is substantial dilution of the self-absorbed line by foreground
emission.  This is the case with Model IV.  Although some dilution is
likely occurring in reality, there is only a fairly narrow range of
conditions which dilute the absorption while leaving 
 the spectral shape undistorted.  Models which require 
substantial dilution also 
need careful fine tuning to produce both 
sufficient emission from the shell and sufficient absorption from the 
core.  

The temperature in the shell is not tightly constrained, and all 
values in the range shown in  Table 1 are acceptable.  The temperature 
must be high enough for the shell to be seen in emission against 
background \hi, but not so high that the optical depth 
towards  the SNR 24.4+0.0 falls significantly below the  $\tau \approx 0.1$ 
measurement of \citet{sanbonmatsu}.

The average volume density of the models will be equal to the true
volume density only if  the filling factor of \hi\ 
is unity.  Our observations do not provide
information on the \hi\ filling factor along the line of sight, but  
 given that the partial pressure of the \hi, $P= nT$,  
is only a few hundred cm$^{-3}$ K, much less than the average 
interstellar pressure (e.g. \cite{hk}) it is likely that 
either the \hi\ filling factor along the line of sight is small 
or the total pressure is dominated by the molecular component of this
cloud. 

The models indicate that the \hi\ cloud 
has a linear extent of 150 pc, a mass in \hi\  of 
$0.5 - 0.8 \times 10^5\  M_{\sun}$, and a peak \hi\ column density
of $5 - 12 \times 10^{20} {\rm cm^{-2}}$, which implies that the extinction 
at visual wavelengths is A$_V = 0.25 - 0.6$ under standard assumptions
about the relationship between gas and dust \citep{spitzer}.
It is interesting that the models give a circular 
shape for the entire cloud, but require a core which is highly elongated along
the Galactic plane.  The transition between warm shell and cold
core thus takes place at  N$_H = 4.6 \times 10^{20}$ cm$^{-2}$ along the
Galactic plane, but at twice this value, $ 9.0 \times 10^{20}$ cm$^{-2}$,
at the maximum z-extent of the core, some 30 pc away from the plane.
Figure 6 shows N$_H$(z) and N$_H(x)$ through the center of 
Model I.  The arrows mark the edge of the cold core.

\section{Anomalous 1720 MHz OH  Emission}

Certain physical conditions can cause a population inversion in transitions of 
 the OH molecule producing  weak emission in the 1720 MHz satellite line 
accompanied by no signal, or absorption, in 
the main 1665 and 1667 MHz lines.   This phenomenon was described
by \citet{turner82}, who distinguished it from the more modest
1720 MHz anomalies observed in dark clouds (\eg\ \citet{turner71}) 
and from the 1720 MHz maser emission observed from compact spots in 
shocks where a SNR is interacting
with a molecular cloud at densities $\sim 10^5 $ cm$^{-3}$ 
(\citet{goss,frail,lockett}, though \citet{yusef} 
argue that a few extended objects also show maser emission).

The Galactic plane in the vicinity  of the \hi\ cloud has been 
observed in all four 18 cm OH lines
by \citet{turner79}, during a general survey, and in the 1720 MHz 
transition  by \citet{frail}, who observed towards 
G27.8+0.6, the supernova remnant which sits atop the radio continuum hole.
Both sets of observations were made with the 140 Foot Telescope
 at an angular resolution of $18\arcmin$.  The channel spacing 
 was 1.13 and 0.85  km~s$^{-1}$ for the Turner and Frail \etal\ 
observations, respectively.
Frail \etal\  published only a cursory summary of their 140 Foot observations
of this region, so we extracted the raw data from the NRAO
 archives and re-reduced them to obtain the values shown 
in Table 2.   The peak line antenna temperature, $V_{LSR}$,
 and line-width, $\Delta v$ (FWHM), were derived by fitting a Gaussian
function to the lines.   The Frail \etal\  data are consistent
with those of Turner in the few directions observed by both.

With the combined Turner and Frail \etal\  data sets there are more than
60 pointings covering the  region around the \hi\ cloud.
Figure 7 shows the positions of 1720 MHz OH 
observations and the boundaries of the \hi\ cloud and inner core 
derived from  Model I.  Circles mark locations 
where 1720 emission was detected at  the velocity of the \hi\ shell
(between 72 and 83 km~s$^{-1}$). 
It is clear that the \hi\ cloud is coincident with a region of anomalous
1720 OH emission  concentrated 
 towards the core of the cloud.  The longitude extent of the
OH emission confirms the conclusion from the 100 meter \hi\  data
that the full extent of the \hi\ cloud core is not much larger than the
region mapped with the 140 Foot.

The width of the 1720 MHz line is  typically 5 to 8~km~s$^{-1}$,
 corresponding to a Doppler temperature $>10^4$ K, which indicates that 
the line width derives entirely from turbulence.  The lines are 
slightly narrower than the \hi\ self-absorption lines  at similar 
positions: the ratio $\Delta v$(\hi)$/ \Delta v(OH) = 1.26 \pm 0.14$
for the seven positions where such a comparison can be made.  
If this difference in $\Delta v$ is interpreted as arising entirely from 
the partial saturation of the \hi\ lines, the ratio implies that 
$\tau \approx 1.5$ for the cool \hi, a value in the range of that 
derived from the cloud models.  

 Figure 8 shows velocity-longitude
and velocity-latitude diagrams for the 1720 MHz emission, \hi\ shell,
and \hii\ regions in the area (from \citet{jay,diffuse}).
The change in velocity with longitude of the OH is entirely 
consistent with the sin($\ell$) 
projection of Galactic rotation on  the LSR for
an object at a constant distance from the Galactic center.
The 1720 MHz OH  emission coincides with the \hi\ cloud, 
although the OH emission appears to be systematically offset 
by $-2.5$   km~s$^{-1}$.  This offset appears in both the \citet{frail}
and the \citet{turner79} OH measurements, and can also be seen in Figure 3.
 We have examined basic observational material 
and can find no error which would create a velocity difference
of this magnitude. 
While it is quite difficult to determine a precise velocity for \hi\ 
self-absorption because the spectral shape of the background emission
can never be known very accurately, all of our attempts to reduce the  
velocity difference between \hi\ and OH 
have been unsuccessful.   There appears to be a 
real velocity difference between 
the \hi\  and the 1720 MHz OH emission from this cloud.

The upper panel of 
Figure 9 shows the integral under the 1720 MHz OH emission, W(OH), in
K km~s$^{-1}$, vs.~the total 
\hi\ column density from Model I, where the model has been smoothed
with an $18\arcmin$ Gaussian to match the angular resolution of the OH data.
The lower panel shows W(OH) plotted against the N$_H$ from the Model I cold
core alone.  The higher degree of correlation of OH and \hi\ in the 
lower panel of Fig.~9, particularly with respect 
to directions lacking OH emission,
indicates that  the  1720 MHz OH emission is correlated only 
with cool \hi, not with the total amount of \hi.

The physical conditions which cause this form of anomalous 1720 MHz
OH emission have not been studied in detail, but \citet{turner82} 
suggested that the kinetic temperature must be confined to a 
fairly narrow range: $15\  {\rm K \leq T \lesssim 50}$  K.  This is 
exactly  the range of temperatures given
by  the \hi\ models for the cloud core, although the
volume densities inferred by Turner for the excitation are  
$\sim 200 \lesssim {\rm n \lesssim 600\  cm^{-2}}$,  orders of magnitude
higher than the average volume density of the \hi.
It is not surprising that there is molecular gas in this cloud, 
for $H_2$ is known to be a significant
component of all interstellar clouds that have N$_H \geq 5 \times 10^{20}$
cm$^{-2}$ \citep{savage}.  What does seem singular, though, is
the correlation between W(OH) and the column density of cold \hi. 
Although the OH molecule is abundant in cool \hi\ clouds 
 \citep{liszt96}, weak, anomalous 1720 MHz emission is much rarer.  
In the inner Galaxy it traces coherent structures many tens of parsecs in 
extent, but it has rarely been found beyond the solar circle
\citep{turner82,turner83}.  The correlation of 1720 MHz OH emission 
 with G28.17+0.05 suggests that this cloud may not be
 unique -- clouds similar to it likely can be found throughout 
the inner Galaxy.

\section{CO}

The \hi\ cloud lies projected on the  edge of a molecular cloud which 
extends to higher longitudes and 
 has been identified in low-resolution  $^{12}$CO spectra 
as one of the largest in the Galaxy with 
$M=10^{6.7} M_{\sun}$ \citep{Myers86}.   However, 
the catalog of \citet{sol87}, derived
from higher angular resolution data, divides the $^{12}$CO emission
 from this part of the Galactic plane 
into many smaller objects, and lists  
five giant molecular clouds toward and around the \hi\ cloud, all with
velocities near that of the \hi.  The $^{13}$CO data of \citet{liszt84},
though taken only in the Galactic plane, show a region of emission
that is centered approximately at the $(l,$v) of the \hi\ cloud with 
 about the same extent in longitude as the cloud core.  
The $^{13}$CO emission has a broad line width similar to that of
the \hi\ and OH.  We believe
that these latter data are the most relevant and give the clearest picture of
the actual association between CO and \hi.
\citet{bania00} have found clouds in which the 
 $^{13}$CO emission tracks \hi\ self-absorption much better than 
the $^{12}$CO lines, which are saturated.  This may also be the 
situation with G28.17+0.05.  Nonetheless, in the absence of complete
 $^{13}$CO maps of the cloud, we  must rely on  $^{12}$CO data for
estimates of its molecular content.  
 
Figure 10 shows a grey-scale map of W(CO): the  power in the $^{12}$CO 
line integrated over the velocity of the \hi\ cloud.
The $^{12}$CO data used here are from the  UMass-Stony Brook survey
of \citet{sand86} which sampled the Galactic plane 
on a $6\arcmin$ grid with $45\arcsec$ angular resolution.  
Circles show directions with associated anomalous 
 1720 OH emission,  stars mark \hii\ regions with the velocity of 
the \hi\ cloud, and 
the outer boundaries of Model I are drawn. There is some concentration
of $^{12}$CO emission towards the \hi\ cloud, but there is 
significant $^{12}$CO emission in this field that appears unrelated to it.

The upper panel of 
Figure 11 shows  W(CO) from Figure 10 between 
longitudes $27\fdg0$ and $29\fdg5$ plotted against  the total
${\rm N_H}$ from Model I evaluated at the location of each $^{12}$CO
spectrum. The general correlation between the model 
\hi\ and the observed $^{12}$CO is quite poor,
and there are directions of  bright $^{12}$CO emission 
where Model I predicts no \hi. Unlike the anomalous 1720 MHz OH emission,
$^{12}$CO is not confined within the boundaries of the \hi\ cloud.
While there may be  a displacement
between the atomic and molecular portions of the cloud, it is 
also likely that some of the  $^{12}$CO emission 
 arises in background molecular gas unassociated with the \hi\ cloud.

The lower panel of Figure 11 shows  W(CO)  compared to the  \hi\ from 
the cold core component of cloud Model I. 
It is apparent that  W(CO) is elevated towards the core relative to 
the general field displayed in 
the upper panel of Figure 11, that it is brightest at the
center of the core, and that no direction towards the core has extremely
low CO emission. 
There still appears to be a significant amount of   $^{12}$CO emission
unrelated to the \hi, perhaps at the level of 30 K km~s$^{-1}$.  We 
adopt the relationship N$_{H_2}$/W($^{12}$CO) $= 2 \times 10^{20}$ cm$^{-2}$ 
\citep{bloemen}, understanding that it may be uncertain by a factor of 
two or more (\eg\ \citet{hunter}).  The total 
 amount of $^{12}$CO towards the cloud core  implies 
a molecular mass M$_{\rm H_2} = 6 \times 10^5 M_{\sun}$.  If a background 
level of  30 K km~s$^{-1}$ is removed,  
M$_{\rm H_2} = 1.4 \times 10^5 M_{\sun}$.

 Figure 12 shows a $^{12}$CO spectrum directly
toward the cloud center at $28\fdg17+0\fdg05$ where the $^{12}$CO data
have been convolved  to the $21\arcmin$ resolution of
the \hi\ data. 
  At the velocity of the \hi\ cloud (marked by the arrow), the $^{12}$CO 
line has a width of $\Delta v \approx 15$ km~s$^{-1}$, about twice
the width of the \hi. 
In contrast to the 1720 MHz OH emission, whose average line width is 
 $75\%$ that of the \hi,  the ratio $\Delta v(CO)/\Delta v(HI)=2.1\pm0.5$
 for nine directions where the comparison can be made.
 Individual $^{12}$CO profiles at the angular resolution
of the UMass-Stony Brook survey show that the broad lines are not 
a result of the convolution, but are intrinsic to the data.
 This trend is general over the region.
Many  $^{12}$CO spectra toward the cloud core have a flattened peak
at the velocity of 
the \hi\ cloud,   suggesting that they may be saturated. 
Saturation of   $^{12}$CO lines from molecular clouds in the
inner Galaxy has also been inferred from  $^{13}$CO observations
\citep{bania00}.

Existing CO observations, especially the  $^{13}$CO data of \citet{liszt84},
show that there is a molecular cloud associated with the \hi\ 
cloud core, but evidence for association with the cloud shell, or 
evidence that this object  is part of a larger interstellar structure, 
remains ambiguous.
It is interesting to contrast the  $^{12}$CO spectrum towards the \hi\ cloud,
which has emission at nearly every velocity up to 120 km~s$^{-1}$, 
with the 1720 MHz OH spectrum, which has emission only at the velocity
of the cloud.  The anomalous 1720 MHz OH emission is 
identifying some special physical conditions present in G28.17+0.05
but  absent in the many other molecular and atomic clouds in this 
direction.

\section{Star Formation}

Information on star-forming regions which might be associated with 
G28.17+0.05 is summarized in Table 3, which includes
 radio recombination line data and tracers of dense molecular regions.
The radio and infrared sources nearest each other 
are given if the identification is not certain.

Almost all of the bright $100\mu m$ and $60\mu m$
sources from the IRAS survey that lie within the \hi\ cloud boundaries 
have been detected in radio recombination lines, but only two 
have velocities near that of the  \hi\ cloud (see Fig.~8). 
There is not enough known about either of these to resolve the 
kinematic distance  ambiguity.
The radio continuum flux density of the two objects detected in ionized gas, 
27.975+0.080 and 28.295-0.377, is 0.2 and 0.5 Jy, respectively, 
[from the 5 GHz survey of \citet{altenhoff}] though the need
for background subtraction introduces some uncertainty.  If this flux
results from optically thin emission (which is likely given the
detection of radio recombination lines at a normal line-to-continuum ratio)
they each require only a single main-sequence B0 star for their
excitation using the Lyman continuum flux calculations of 
\citet{vacca}.  Of particular interest are two IR and sub-mm sources
associated with the radio source  28.332+0.060.  One has a
bolometric luminosity typical of a pre-main sequence star, while the other
is a candidate for classification as a class 0 protostar \citep{carey00}.  

Maser emission from the  6.7 GHz transition of ${\rm CH_3OH}$ is thought
to trace sites of the formation of massive stars \citep{menten}, and
its detection towards several IR sources indicates 
that there might be 
ultra-compact \hii\ regions \citep{wood} associated with
the cloud, although there is always the
possibility of confusion with a background object at the far
kinematic distance.  There is a bright infrared source projected near
the  center of the \hi\ cloud at $28\fdg20-0\fdg05$ 
\citep{caswell,szymc}, 
but its ${\rm CH_3OH}$ emission covers 94-104 km~s$^{-1}$ 
and its CS(2-1) emission is at 97.4 km~s$^{-1}$, placing 
it well behind the \hi\ cloud.
\citet{ellingsen} have argued that many 6.7 GHz ${\rm CH_3OH}$
masers are not associated with IRAS sources, so there may be as yet
undetected young objects in the cloud.

\citet{Myers86} have studied the star formation rate in molecular
clouds in the inner Galactic plane and list a molecular cloud at 
$l=29\deg, b= +80\ $ km~s$^{-1}$, as one of the most massive in their
survey, but with a negligible number of \hii\ regions and
little far-Infrared emission. Higher resolution molecular observations 
suggest that what Myers \etal\ reported as 
 one cloud is probably several (\S 6), but the new data 
 summarized here  do not contradict 
the general conclusion of Myers \etal\  that this cloud has relatively
little star formation compared to most large clouds in the inner Galaxy 
(see also \cite{carey98}).

\section{Comparison with Maddalena's Cloud}

There is a large molecular cloud in the outer Galaxy, 
G216-2.5, often called Maddalena's Cloud,  which has unusual 
 properties: a relatively large line width (in $^{12}$CO) of 
$\Delta v = 8.5$  km~s$^{-1}$, a size of $50 \times 150$ pc, and 
a molecular hydrogen mass derived from $^{12}$CO observations 
(assuming  N$_{H_2}$/W(CO) of $2 \times 10^{20}$ cm$^{-2}$) 
of   $2 \times 10^5$  M$_{\sun}$.  It is associated
with an \hi\ cloud that has a mass M$_{\rm H~I} \approx 5 \times 10^4 M_{\sun}$
and a size  $\approx 100 \times 300$ pc
\citep{mad85,wilmad}.
The kinetic temperature of the \hi\ is unknown. At its peak, 
the \hi\ has  T$_b \approx 50$ K but some of this
is probably unrelated emission.  The cloud is not seen in \hi\ 
self-absorption, but the level of background \hi\ emission is 
uncertain, and may not be bright enough to produce the effect.  
This cloud shows no evidence for current star formation, nor 
is it a source of anomalous 1720 MHz OH emission \citep{turner79}.
Williams \& Maddalena have proposed that the \hi\ arises from 
photodissociation of part of the molecular cloud by an O9.5 star 
located 50 pc from the  molecular cloud edge.

Maddalena's cloud is in some ways  similar to G28.17+0.05, but in
important ways distinctly different.
The main similarities  are in the large, turbulent 
line width,  its size, and the association of \hi\ and molecules 
in a single object. This is shown in Table 4.
Differences are that Maddalena's cloud has no anomalous 1720 MHz OH emission,
absolutely no  evidence of star formation, and no evidence 
 for a cold \hi\ core.

\section{The Nature of G28.17+0.05}

\subsection{Mass}

A strong lower limit to the $H_2$ mass of G28.17+0.05 is
 $1.4 \times 10^5$ M$_{\sun}$ (\S 6).  Adding the \hi\ mass estimated from 
Model I and scaling by a factor of 1.36 to account for heavy elements 
yields a total  cloud
mass of no less than $2.4 \times 10^5$  M$_{\sun}$.
  The virial mass of the cloud,
 derived from its size and linewidth, is in the range
$3 - 8 \times 10^5 M_{\sun}$, depending on assumptions about
the cloud structure \citep{mckee}.  The virial mass 
 is so close to the observed mass that, 
given the rather substantial uncertainties, it is plausible that 
the cloud is gravitationally bound.  As discussed below,
this makes G29.17+0.05 unusual because the \hi\ on the edges of most
molecular clouds is not gravitationally bound.

\subsection{Relationship Between Atomic and Molecular Components}

Many molecular clouds have 
significant amounts of atomic gas: either cold \hi\ in a
cloud's core, an extensive warm \hi\ halo, or some combination of both
\citep{andersson91,feldt,minh}.
Every   warm \hi\ halo found thus far is apparently transitory, 
 either because
the object it surrounds is not long lived, or the \hi\ is too turbulent to 
be confined by the gravitational field  of the cloud (\eg\
\citet{andersson93,minh}).    A study of several molecular clouds 
found that the line width of  \hi\ halos was 5 times larger than 
the line width of the molecules \citep{andersson91}. In some cases the \hi\ 
halo is thought to be material dissociated from a molecular cloud by
nearby stars \citep{kuchar,wilmad}. \citet{allen97} suggest that all 
of the bright \hi\ in the spiral arms of the galaxy M81 arises from  
dissociated  molecular gas.

G28.17+0.05 thus appears anomalous, as it seems sufficiently massive
to be gravitationally bound despite its large line width.  
It is also anomalous in having a similar line width for \hi\ and 
OH. G28.17+0.05 is unlike the transitory objects noted above.

\subsection{Atomic Fraction}

	\citet{turner97} has reviewed the formation of molecular clouds
from diffuse interstellar gas.  The transition from a fully atomic to 
a fully molecular hydrogen cloud can be abrupt once sufficient
shielding from ambient UV is available.  The column density of \hi\ 
through the G28.17+0.05 core is estimated to be 
$12 \times 10^{20}$ cm$^{-2}$ 
(Model I) so dust associated with the \hi\ provides an extinction of 
only A$_V < 1$ magnitude through the 
cloud core.  Of course, the molecular observations imply that 
most of the cloud is in the molecular state, so the total  A$_V$ must
be substantially higher.

The models give a mass in cold \hi\ in the cloud core of
$0.8 - 3 \times 10^4$  M$_{\sun}$ which is at most $6\% - 20\%$ of the 
mass in H$_2$ of the core. 
In most molecular clouds, the column density of cold \hi\ is only 
a few $10^{18}$ cm$^{-2}$ and is $\lesssim 1 \% $ of the total gas mass 
(\eg\ \citet{mccutch,feldt}), 
though it can reach a few $10^{20}$  cm$^{-2}$ in some regions 
\citep{gibson,bania00}.
Chemical models of molecular clouds predict a residual \hi\ content of 
just a few percent unless the external radiation field is quite high 
and much of the molecular cloud is dissociated 
(Wolfire, Hollenbach \& Tielens 1993).  

\subsection{Significance of the Line Width}

Turbulence appears to be the dominant source of pressure in the \hi,
and as discussed in \S9.1, the turbulent pressure is approximately
that needed to support the cloud against gravitational collapse.
The uncommonly large line width $\sim 7 $ km s$^{-1}$ 
in  \hi, OH, and $^{13}$CO, and  
the near-equivalence of the line width in the core and shell despite
their differing physical temperature, suggests that turbulent motions 
of the same magnitude are present in all parts of the cloud.  

Turbulence is expected to dissipate on the timescale of the turnover
of a turbulent eddy, which is on the order of the turbulent velocity 
divided by the size of the cloud  \citep{turb}.  For G28.17+0.05 
the time scales are less than a few $10^7$ years.  This suggests
that the turbulence in the cloud has been created recently, which 
might be the case (a) if the cloud is in the process of coalescing from 
more distributed material; (b)  if it is being shocked ({\it e.g.},
passing through a spiral arm); (c) if there is a turbulent mixing layer
between the cold core and the shell; (d)   if it is being stirred by 
the products of star formation.  It is likely that this cloud has 
undergone some recent perturbation and is not in equilibrium, 
 consistent with the unusually large fraction of its 
matter which is in the atomic state.  

\subsection{Analogs to G28.17+0.05 in Other Galaxies}

There must be analogs to G28.17+0.05 in other galaxies, so it is 
instructive to consider how it would appear when viewed from above,
instead of being silhouetted against emission from the Galactic plane.
The most important difference is that gas above and below it in the 
Galaxy is expected to have N$_H \approx 3 \times 10^{20}$  cm$^{-2}$
in each half-disk \citep{dl90} which would produce a
background  T$_b < 10$ K.  
The cloud would appear completely in emission against
this weak background, and have a maximum T$_b$ of 50 - 60 K, and an
extent of 150 pc in azimuth around the galaxy, i.e. in the direction 
approximately parallel to the spiral arms.  The maximum column 
density would be about $10^{21}$  cm$^{-2}$.  The cloud's shape 
in the radial direction, perpendicular to the spiral arms, cannot
be determined from our data.

\citet{braun} describes the \hi\ in nearby
galaxies as being concentrated in a network of emission features with 
a narrowest dimension of 150 pc and \hi\ brightness temperatures of 
80-200 K.  The core of G28.17+0.05 cannot be hotter than 50 K, but this
is within the range of observed values \citep{braun}.  
It is possible that G28.17+0.05 is a Galactic analog of this
extragalactic phenomenon.

\subsection{A New Giant Molecular Cloud in the Making?}

The transition between a cloud consisting of \hi\ and one which is
mostly  H$_2$ depends on the local pressure to
a very high power \citep{elmegreen93} such that the interaction of an \hi\ 
cloud with a spiral density wave can be expected to convert it 
to H$_2$, and the \hi/H$_2$ ratio is expected to be a strong function
of distance from the Galactic plane \citep{elmegreen93,combes}.
The kinematic distance adopted for the \hi\ cloud is 5 kpc, which
corresponds to a distance from the Galactic center, R,  of 4.8 kpc.  This
is near the location  of the Scutum spiral
arm, which is seen tangent to the line of sight near longitude $30\deg$
implying R=4.25 kpc.  This raises the possibility that 
the \hi\ cloud is interacting with the potential minimum of the arm and
that the distinctive properties of the cloud come from the fact that it is
being observed as it enters a spiral shock and begins a phase transition
from  atomic to  molecular gas.

 The general scenario for interaction of a cloud with a spiral
shock is the formation of molecules from predominantly atomic gas, 
a velocity discontinuity at the shock, and subsequent star formation leading
eventually to the disruption of the molecular cloud
 \citep{allen86,elmegreen95}.  
The cloud G28.17+0.05 would fit into this sequence if 
 it is just now in the process of forming molecules and 
initiating star formation.  This would account for its relatively
high fraction of \hi\ and its relatively low star formation rate. 
 In the inner Galaxy the
spiral pattern speed is less than Galactic rotation, so gas 
overtakes the spiral potential.  The spiral shock would slow 
gas reducing its apparent rotational velocity. In the first quadrant of 
Galactic longitude, this would result in a blue-shift of 
shocked gas relative to pre-shocked gas.  
This provides a natural explanation for the observed 2.5 km~s$^{-1}$  
difference in velocity between the \hi\ and the OH,  if  the anomalous 1720 MHz
OH emission comes from shocked gas, and the \hi\ from  pre-shocked gas.
The total velocity difference may actually 
be higher than this value because of projection effects.
If this suggestion is correct, and the mechanism is general, 
 then a similar cloud in the fourth quadrant of Galactic 
longitude will have anomalous 1720 MHz 
 OH emission red-shifted with respect to the associated \hi\  
because of the different projection of Galactic rotation in that quadrant.

One puzzling feature of the  anomalous 1720 OH emission is 
that it is seen in many directions in the inner Galactic
plane, but must be quite localized because it 
is rarely seen at more than one velocity and requires special 
excitation conditions \citep{turner82,turner83}.  If this transition is
excited in a spiral shock both properties could be understood, for 
a spiral shock is both widespread and localized.  \citet{turner82} 
noted that the 1720 MHz transition tracks spiral arms in the inner Galaxy 
much more clearly than either $^{12}$CO or H$_2$CO, a 
behavior that would be expected in this scenario.

\section{Concluding Comments}

The \hi\ cloud at $28\fdg17+0\fdg05$ is unique in a number of respects.
First, it is very large and unusually well defined for a Galactic \hi\
cloud.  Its extent of  150  pc and its \hi\ mass approaching
$10^5$ M$_{\sun}$ are at the extreme range for Galactic \hi\ clouds, 
 quite unlike the usual objects detected in \hi\ self-absorption
(\eg\ \citet{knapp,gibson}).

Second, the cloud has a very distinctive
temperature structure, with a cool core at 25-45 K 
and a warmer shell at 150 - 400 K.  While some small
\hi\ clouds are known to have a similar temperature structure 
(\eg\ \citet{vanderwerf,feldt}), 
nothing on so large a scale or with such a steep gradient has been
observed before.  Moreover, the line width of 7 km~s$^{-1}$, even in the 
cold gas, is unusually large for 
cool \hi\ clouds and indicates that the cloud is quite turbulent 
(cf.~G28.17+0.05 with the clouds in  \citep{knapp} and \citep{gibson}).

Third, the cloud core, but not the outer shell, is associated with
 anomalously-excited 1720 MHz OH emission.
This in itself shows that G28.17+0.05 is not a typical molecular cloud. 
The excitation mechanism for this emission has not been 
studied in detail, but conditions which produce it must be 
 fairly widespread in the inner Galaxy.  The OH lines have a  
width consistent with that of the \hi, but appear to be blue-shifted
by about 2.5 km~s$^{-1}$.  The cloud core, but not
the shell, also seems to correlate with  emission in the $^{12}$CO
and $^{13}$CO lines.  The brightest molecular emission occurs in 
directions with the largest amount of cold \hi. 
The estimated H$_2$ mass associated with the
cloud core, though  uncertain, may be as low as $1.4 \times 10^5$
M$_{\sun}$.  There is some star formation in the cloud, but
relatively little compared with other large molecular clouds.

The overall picture we derive is of a turbulent atomic and molecular cloud. 
Some part of it is in a unusual state of excitation producing the 
anomalous 1720 MHz OH emission.  There is 
some star formation.  We speculate that this is an 
object just entering the Scutum Spiral Arm and initiating molecule
formation and star formation.  The velocity difference between OH and
\hi\ results from their being located in different parts of a shocked
cloud.  If this explanation is correct, and shocks are  a general feature of 
clouds of this type, then a similar cloud in the fourth Galactic quadrant
should show anomalous 1720 MHz OH emission 
that is red-shifted with respect to its associated cool \hi.  

The absence of radio continuum emission near the cloud -- the feature
that first drew our attention to this part of the Galactic plane
 -- results in part from the relatively low star formation rate 
in the cloud, but is probably just a coincidence, for the site line 
toward G28.17+0.05  passes near many areas of significant star formation.
The emission from 
a single foreground or background \hii\ region or SNR would easily destroy 
the appearance of a hole in the radio continuum.
In view of all the other evidence about this cloud,
we consider the existence of radio continuum hole of minor importance, 
and most likely a simple coincidence.

There should be many more clouds like G28.17+0.05 waiting to be
discovered in the inner Galaxy.  This object was 
found only because it lay in a region that was 
completely mapped in \hi\ 
at an angular resolution of $21\arcmin$, sufficient to reveal the
cloud structure.  It is not widely appreciated that 
the majority of the Galactic plane within 
$50\deg$ of the Galactic Center has not yet been observed at 
this high an angular resolution.  
Existing data (\eg\ the Leiden-Dwingeloo \hi\ survey of
\citet{hartman}), while more than adequate for the study of 
the large-scale structure of the Galaxy, do not have 
the angular resolution and completeness
of sampling necessary to detect this cloud, even as large as it is!

The statistics of anomalous 1720 MHz OH emission also suggest that clouds
in this thermodynamic state must be common.  \citet{turner82} found $>50$
 highly extended objects in the region $337\deg \leq \ell \leq 50\deg$, 
and the OH observations of \citet{frail} presented in Table 2 
contain additional directions with this emission.  
It is reasonable to expect
that there are at least 100 large clouds like the one studied here, 
with anomalous 1720 MHz OH emission 
and an extensive cold atomic core, within $50\deg$ of the Galactic center.
The covering factor of these clouds in the inner Galactic plane  
may be $\sim 1$.
Further observations of this cloud at higher angular resolution in 
\hi\ emission and absorption against background continuum sources, 
and in molecular species, particularly $^{13}$CO and CS, 
 should give especially good insights into
the evolution of molecular clouds in the Galaxy.  
The importance of anomalous 
1720 MHz OH emission as a tracer of Galactic structure
has not been appreciated.  New studies of 
excitation mechanisms for this transition would be especially welcome.

\acknowledgments
We thank  Dana Balser, Tom Bania, John Dickey, 
Dale Frail, Steve Gibson, Harvey Liszt, Ron Maddalena 
and especially Barry Turner for useful discussions.  We also thank 
 Mark Heyer for providing the UMass-Stony Brook $^{12}$CO survey 
and  H.S. Liszt for providing the 100 meter \hi\ data.
The research of J.A.L. at NRAO was supported by the NSF Research 
Experiences for Undergraduates program.  The AIPS++ (Astronomical Information
Processing System) is a product of the AIPS++ Consortium.  AIPS++ 
is freely available for use under the Gnu Public License.  Further
information may be obtained from \anchor{http://aips2.nrao.edu}{http://aips2.nrao.edu}.

\begin{deluxetable}{ccccccclccccc}
\tablecolumns{13}
\tablenum{1}
\tablewidth{0pt}
\tablecaption{Models for the \hi\ Cloud.}
\tablehead{
\colhead{} &  \colhead{Shell} & & \multicolumn{3}{c}{Core} & &  
\multicolumn{2}{c}{T$_{ex}$} & \colhead{} & \colhead{} & \colhead{} &
\colhead{}  \\
\cline{2-2} \cline{4-6} \cline{8-9} \\
\colhead{} & \colhead{$r_x$} & &
\colhead{$r_x$}  & \colhead{$r_y$} & \colhead{$r_z$} & & \colhead{Core} &
\colhead{Shell} &
\colhead{n$_0$} & \colhead{n$_r$} &  \colhead{z$_0$} &
\colhead{M$_{\rm H~I}$}  \\
\colhead{Model} & \colhead{(pc)} & &
\colhead{(pc)}  & \colhead{(pc)} & \colhead{(pc)} & & \colhead{(K)} &
\colhead{(K)} &
\colhead{(cm$^{-3}$)} & \colhead{(cm$^{-3}$)} &  \colhead{(pc)} &
\colhead{($10^4 M_{\sun}$)}  \\
\colhead{(1)} &\colhead{(2)} & & \colhead{(3)} &\colhead{(4)} &
\colhead{(5)} & & \colhead{(6)} &\colhead{(7)} &\colhead{(8)} &
\colhead{(9)} &\colhead{(10)} &\colhead{(11)} 
}
\startdata
I & 75 & &  56 & 75 & 30 & & 43.5 & 350 & 4.0 & 1.0 & 4 & 
$7.6 $ \\
II & 80 & & 60 & 75 & 26 & & 40 & 400 & 4.0 & 0.6 & 7 &
$6.8 $ \\
III & 80 & & 56 & 75 & 26 & & 42 & 150 & 4.0 & 1.0 & 7  &
$8.2 $ \\
IV & 75 & & 49 & 52 & 30 & & 25 & 350 & 1.0 & 1.1 & 4 & 
$4.7 $ \\
\enddata
\tablecomments{The semi-major axes of the shell 
$r_y = r_z = 75$ pc for all models.}
\end{deluxetable}

\begin{deluxetable}{rcccc}
\tablecolumns{5}
\tablenum{2}
\tablewidth{0pt}
\tablecaption{1720~MHz OH Emission}
\tablehead{
\colhead{$l,b$}  & \colhead{rms (mK)} & \colhead{T$_{pk}$ (mK)}
& \colhead{$\Delta v ({\rm ~km~s^{-1}})$}  &
\colhead{$V_{LSR} ({\rm ~km~s^{-1}})$} \\
\colhead{(1)} &\colhead{(2)} &\colhead{(3)} &\colhead{(4)} &\colhead{(5)} \\
} 
\startdata
    $27\fdg161+0\fdg737$ &    32 &    \nodata  &  \nodata  & \nodata \\
    $27.276+0.516$ &   30  &     \nodata  &  \nodata  & \nodata \\
    $27.391+0.296$ &   40  &            92  &    5.1 &    75.3  \\
    $27.391+0.296$ &   40  &             142 &     5.4 &    94.5 \\
    $27.506+0.075$ &   35  &            180  &   6.7  &   74.7 \\
    $27.506+0.075$ &   35  &            110  &   4.8  &   96.8 \\
    $27.382+0.852$ &   30  &     \nodata  &  \nodata  & \nodata \\
    $27.498+0.632$ &   37  &     \nodata  &  \nodata  & \nodata \\
    $27.613+0.411$ &   34  &            127  &   4.3  &   74.0  \\
    $27.728+0.190$ &   32  &            245  &   5.8  &   75.7  \\
    $27.728+0.190$ &   32  &             59  &    5.5 &    95.5  \\
    $27.604+0.969$ &   30  &      \nodata  &  \nodata  & \nodata \\
    $27.719+0.748$ &   36  &      \nodata  &  \nodata  & \nodata \\
    $27.835+0.527$ &   35  &            82   &   4.7  &   74.7  \\
    $27.950+0.306$ &   35  &            275  &   5.9  &   75.5  \\
    $27.950+0.306$ &   35  &            72   &   5.2  &   97.6  \\
    $27.826+1.084$ &   33  &      \nodata  &  \nodata  & \nodata \\
    $27.941+0.863$ &   34  &      \nodata  &  \nodata  & \nodata \\
    $28.057+0.642$ &   33  &      \nodata  &  \nodata  & \nodata \\
    $28.171+0.421$ &   35  &            153   &  6.9   &  74.8 \\
\enddata
\end{deluxetable}

\begin{deluxetable}{ccccc}
\tablecolumns{5}
\tablenum{3}
\tablewidth{0pt}
\tablecaption{Star Forming Regions Toward G28.17+0.05}.
\tablehead{
\colhead{$l,b (radio)$} & \colhead{$V_{\rm H~II} ({\rm ~km~s^{-1}})$} & 
\colhead{$V_{mol} ({\rm ~km~s^{-1}})$} & 
\colhead{$l,b (IR)$} & \colhead{IRAS name} \\
\colhead{(1)} &\colhead{(2)} &\colhead{(3)} &\colhead{(4)} &\colhead{(5)} 
} 
\startdata
$27\fdg975+0\fdg080$ & $83.2\pm2.4$ & $75.0$ & $27\fdg975+0\fdg093$ & $18394-0425$ \\
$28.295-0.377$ & $75.6\pm4.2$ & $79-93$ &  $28.302-0.355$ & $18416-0420$ \\
$28.332+0.060$ & \nodata    & $67-81$   & $28.393+0.086$ & $18402-0403$ \\
\enddata
\tablecomments{The molecular velocity V$_{mol}$
for $27.975+0.080$ is for the peak
of the CS(2-1) emission; for the other objects it is the range of
 CH$_3$OH emission.  The data are from 
\cite{jay,diffuse,caswell,szymc,Bronfman}.}
\end{deluxetable}

\begin{deluxetable}{cccrc}
\tablecolumns{5}
\tablenum{4}
\tablewidth{0pt}
\tablecaption{Comparison of G28.17+0.05 and Maddalena's Cloud (G216-2.5)}.
\tablehead{
\colhead{Object} & \colhead{$\Delta v ({\rm ~km~s^{-1}})$} & 
\colhead{M$_{\rm H~I}$ (M$_{\sun}$)} &
\colhead{M$_{\rm H_2}$ (M$_{\sun}$)} &
\colhead{Extent (pc)} \\
\colhead{(1)} &\colhead{(2)} &\colhead{(3)} &\colhead{(4)} &\colhead{(5)} 
} 
\startdata
G28.17+0.05 	  &  $7.0$ & $8 \times 10^4$ &  $ \sim 1 \times 10^5$ & 
$150 \times 150$  \\
G216-2.5 &  $8.5$ & $5 \times 10^4$  &  $ 2 \times 10^5$ & 
$300 \times 100$ \\
\enddata
\tablecomments{ Line-width for G28.2+0.1 is from \hi; for Maddalena's cloud it
is from $^{12}$CO.  The masses account only for Hydrogen, and a conversion
factor of $2.0 \times 10^{20}$ was assumed between N$_{\rm H_2}$ and 
W($^{12}$CO).  Properties of G28.17+0.05 are from Model I.}
\end{deluxetable}


\begin{figure}
\plotone{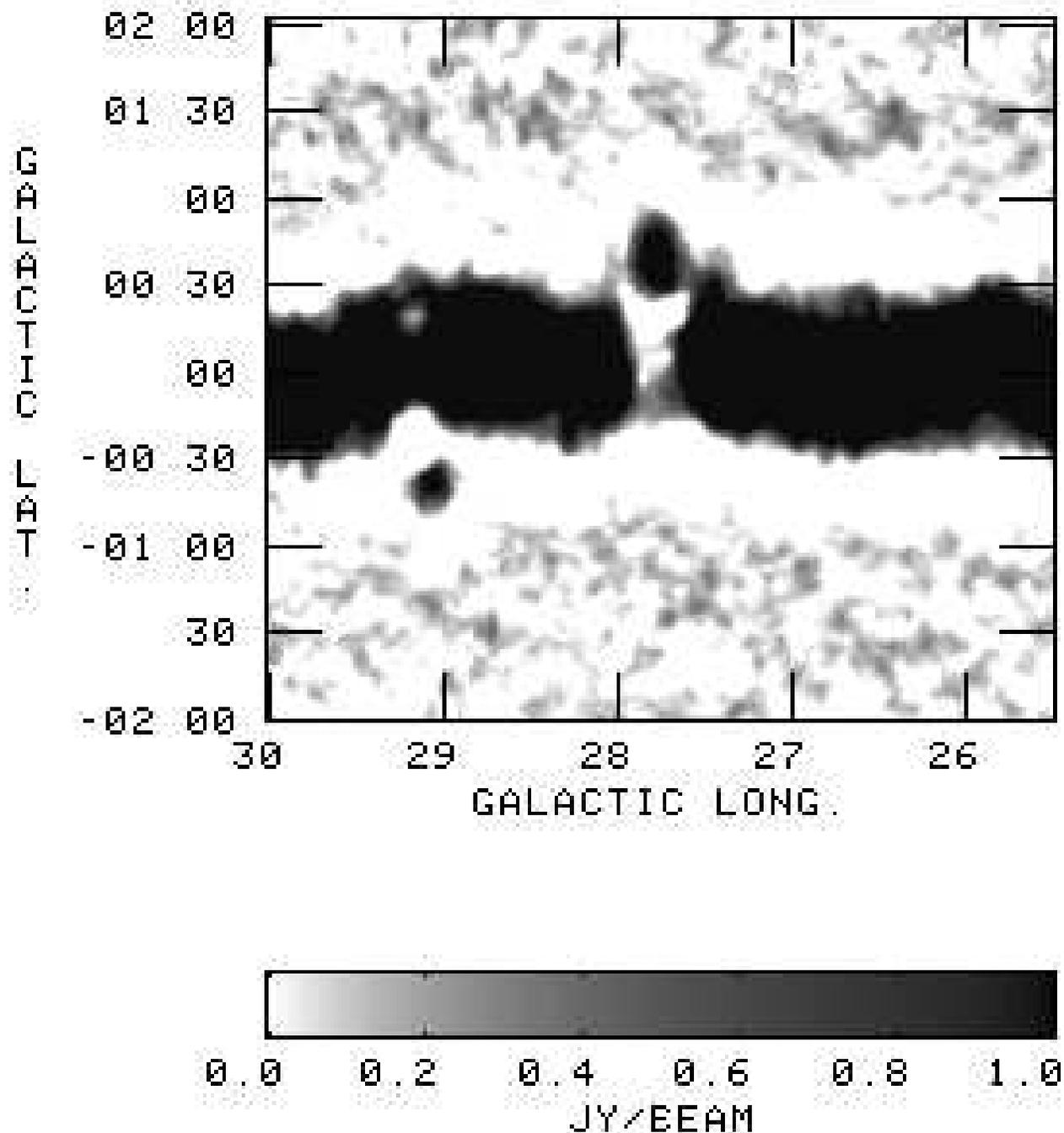}
\caption{The region around $l=28\deg$ from the GPA 8.35 GHz continuum 
survey \citep{gpa}.  This region is shown as a negative image 
to highlight 
the hole in the Galactic continuum.  Spatial filtering suppresses emission
more extended than $1\deg$ in latitude and accounts for the
empty areas parallel to the Galactic equator.  
The flux scale is indicated by the bar graph below the image.}
\end{figure}

\begin{figure}
\plotone{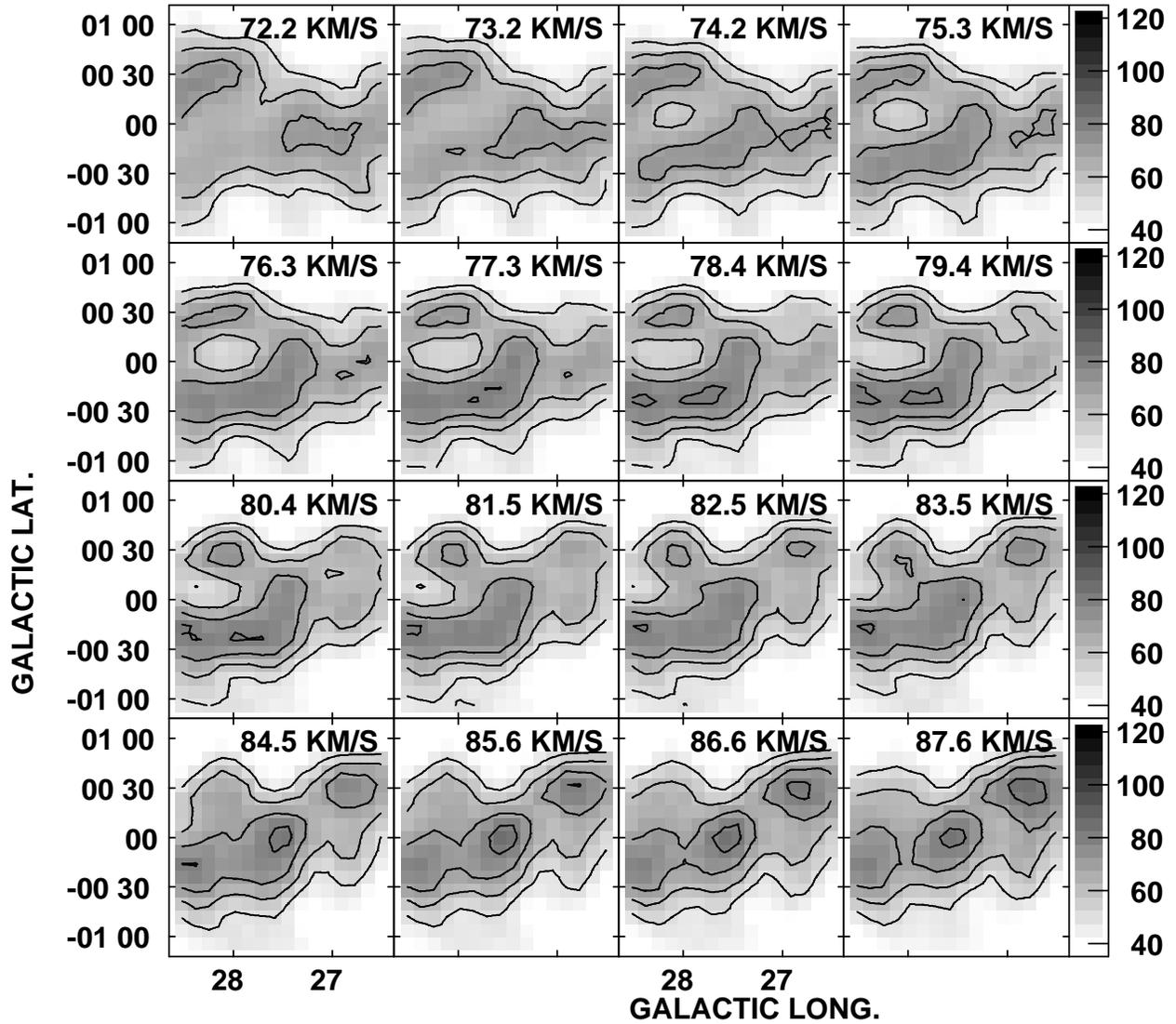}
\caption{
Channel maps of \hi\ brightness temperature vs.~Galactic longitude and
latitude showing the \hi\ self-absorption 
feature centered near longitude $28\deg$ at LSR velocities 
of 74-83 km~s$^{-1}$. The depression due to self absorption is surrounded
by a shell of \hi\ emission.  
The bar graph on the right of the figure shows
the gray-scale flux scale.  The contour levels are at 50, 60, 70 and 80 K.}
\end{figure}

\begin{figure}
\plotone{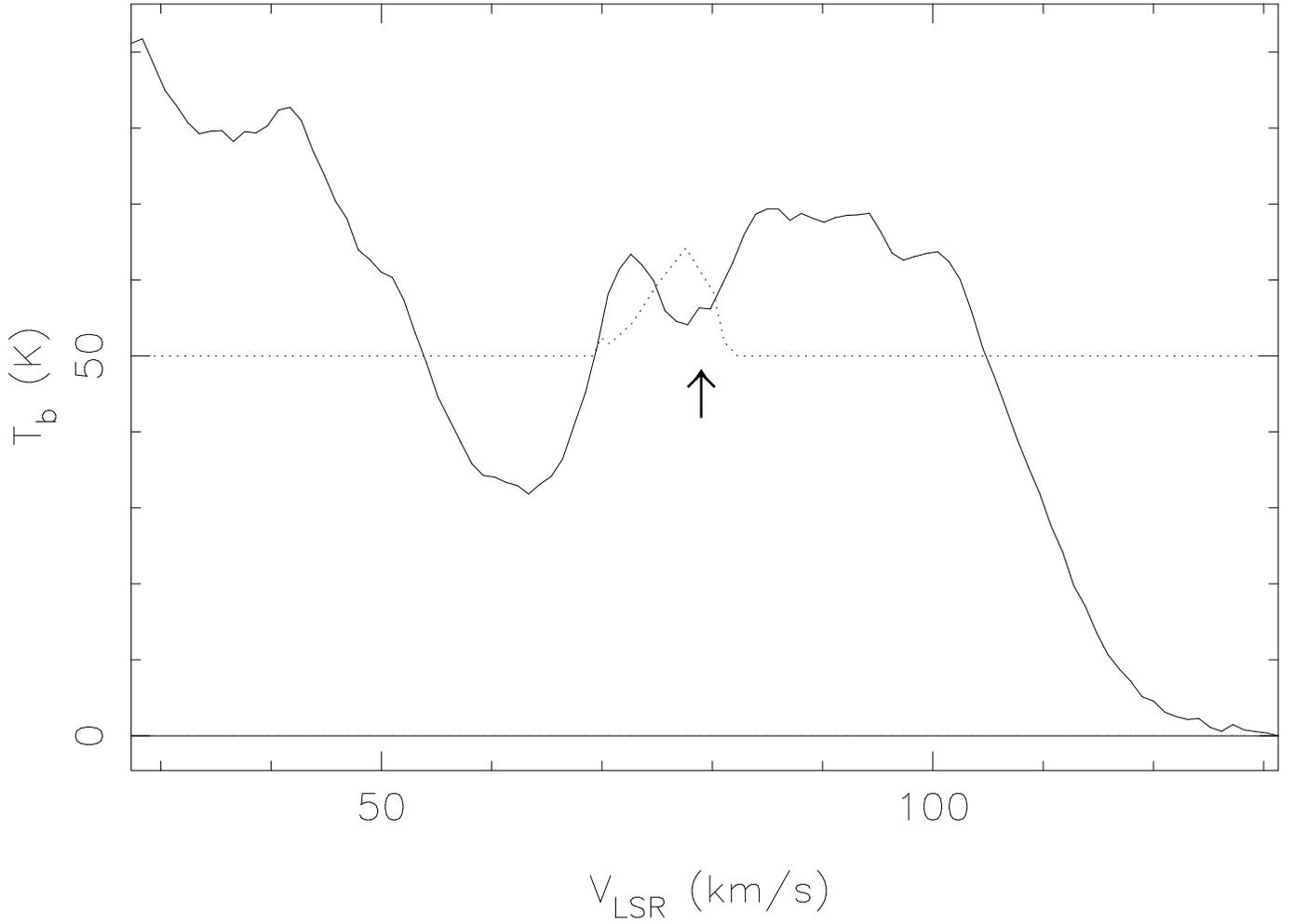}
\caption{
The solid line is a portion of the 21cm \hi\ spectrum towards 
$28\fdg00+0\fdg00$.  The arrow marks the self-absorbed feature.
The dashed line shows the  1720 MHz OH emission line in a nearby
direction (copied by hand from a Figure in \cite{turner79}), 
scaled in intensity  and offset for easier comparison with the \hi. }
\end{figure}

\begin{figure}
\plotone{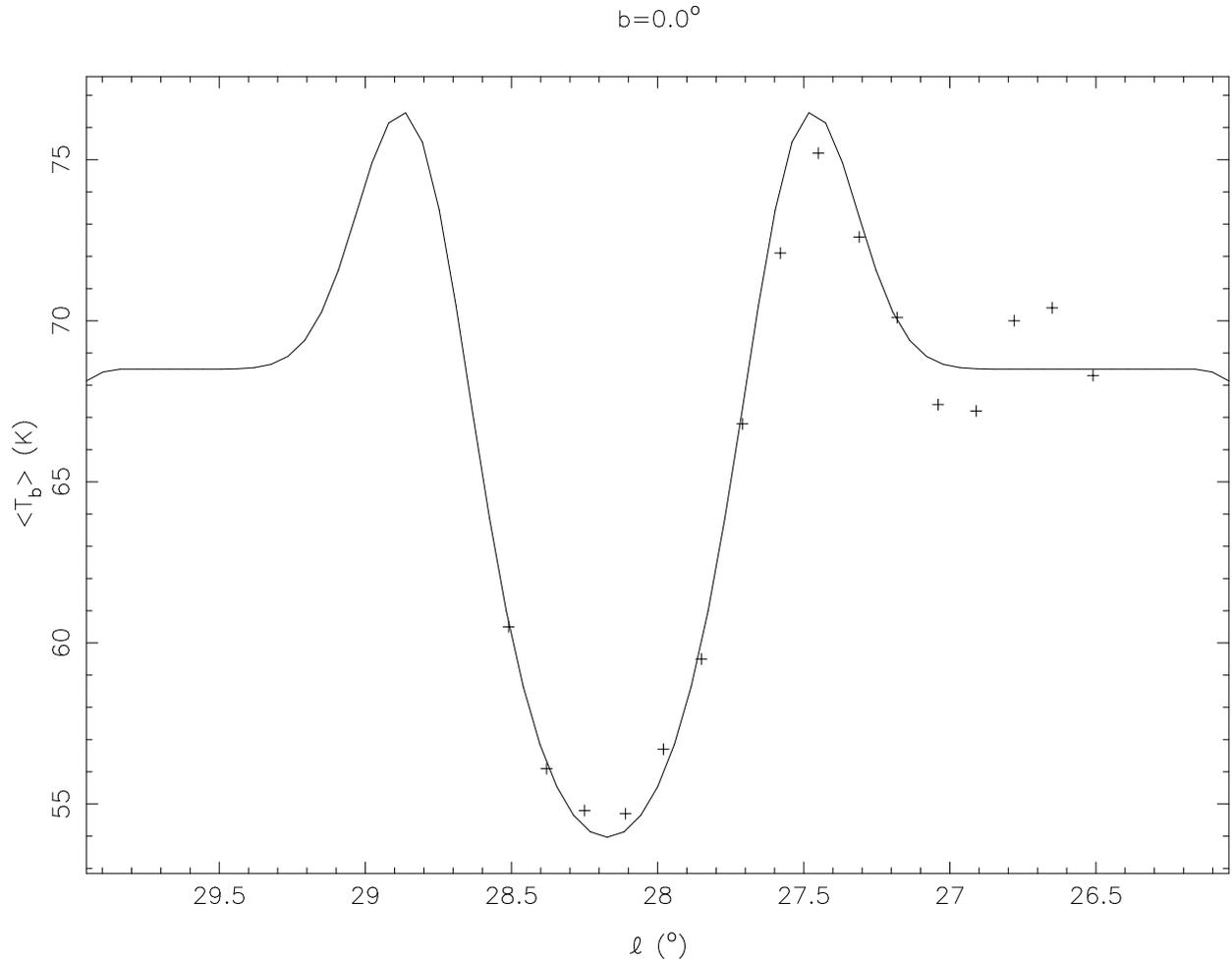}
\caption{
Data points show the \hi\ brightness temperature from 140 Foot 
observations at the velocity of 
the cloud along a strip at $b=0\deg$.  The solid line is the
result of Model I.  The cloud shell is brightest at $\ell = 27\fdg5$ 
while self-absorption is greatest at $\ell = 28\fdg2$. The 100 meter \hi\ 
data confirm that the model is generally correct at the higher
longitudes.
 }
\end{figure}

\begin{figure}
\plotone{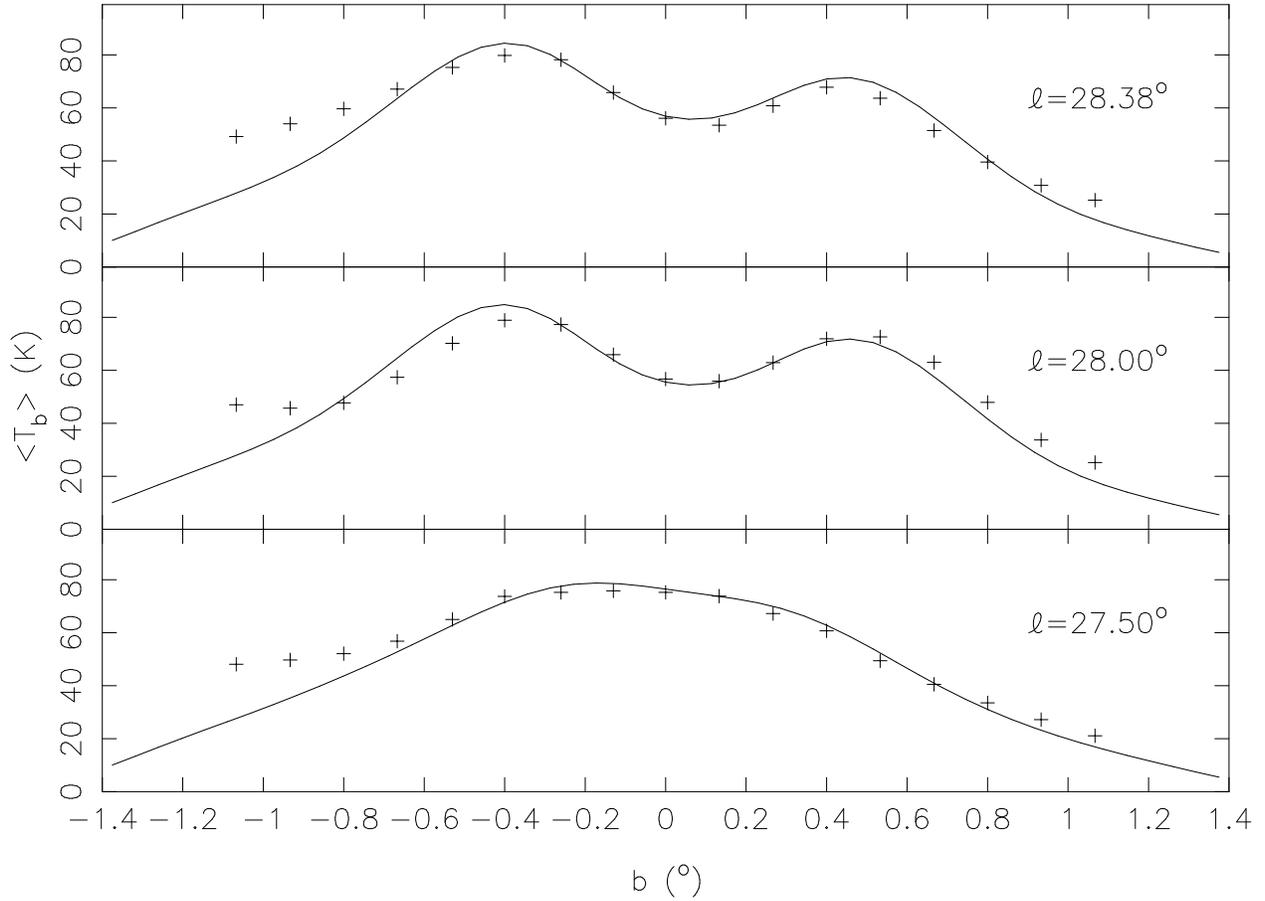}
\caption{The data points are measured \hi\ brightness temperatures averaged
over 7 km~s$^{-1}$ around the central velocity of the \hi\ cloud, 
and the solid line is 
the result from Model I.  The data are taken along cuts through 
 the Galactic plane at constant longitude.
}
\end{figure}

\begin{figure}
\plotone{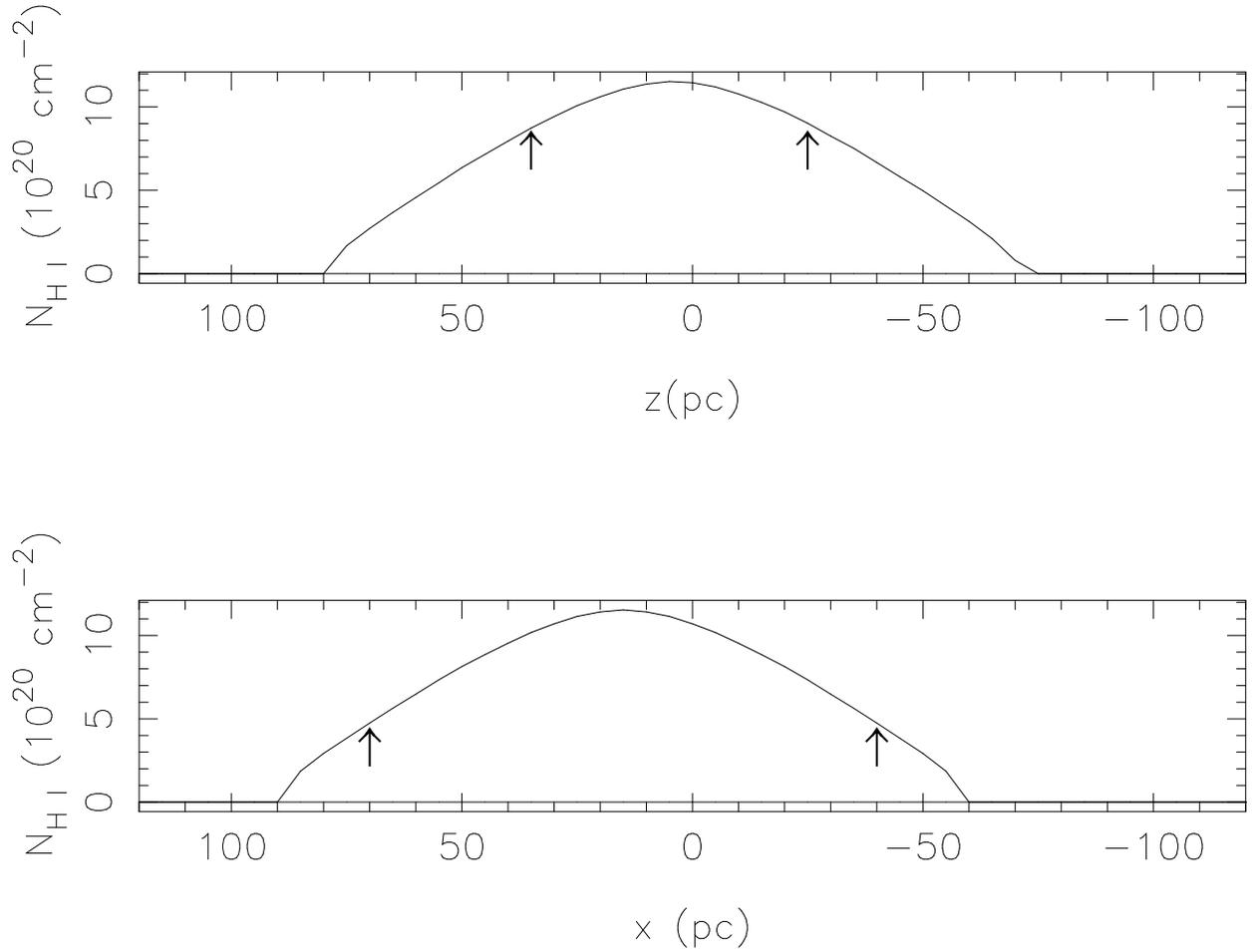}
\caption{Profiles of N$_{\rm HI}$ from Model I through the cloud center along
axes perpendicular to the Galactic plane (upper panel) and parallel to
the plane (lower panel).  Arrows mark the transition between the cold core
and the warm shell.  The core is elongated along to the Galactic plane.  
The coordinate system is centered at $\ell,b = 28\fdg0+0\fdg0$ and 
the cloud is assumed to be at a distance of 5.0 kpc.
}
\end{figure}

\begin{figure}
\plotone{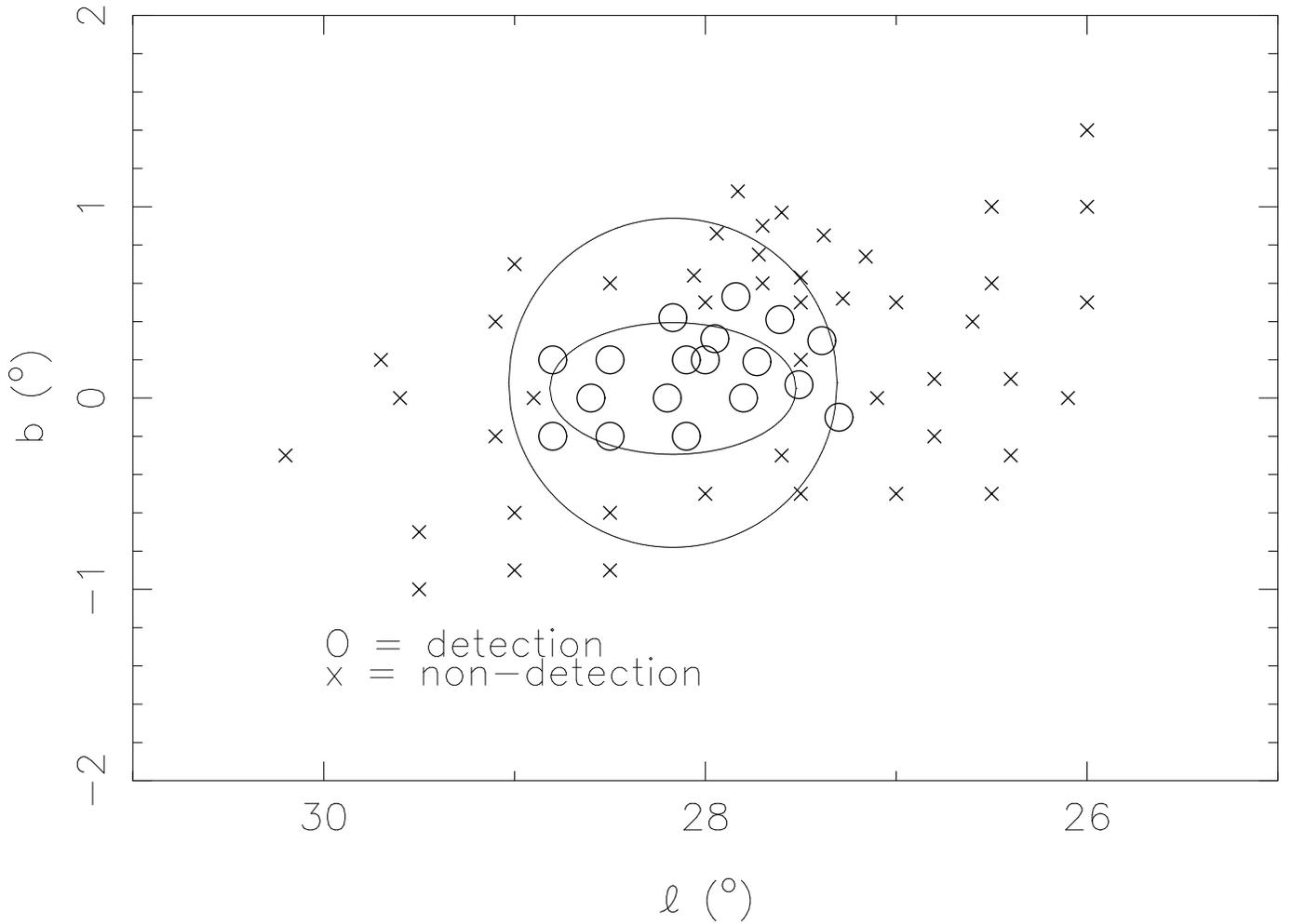}
\caption{Small circles mark locations where anomalous 1720 MHz OH
emission has been detected at the velocity of the \hi\ cloud, and 
crosses mark locations of observation without detection.  The 
boundries of the \hi\ cloud and 
its cold core are shown by the large circle and the ellipse.
}
\end{figure}

\begin{figure}
\epsscale{1.5}
\plottwo{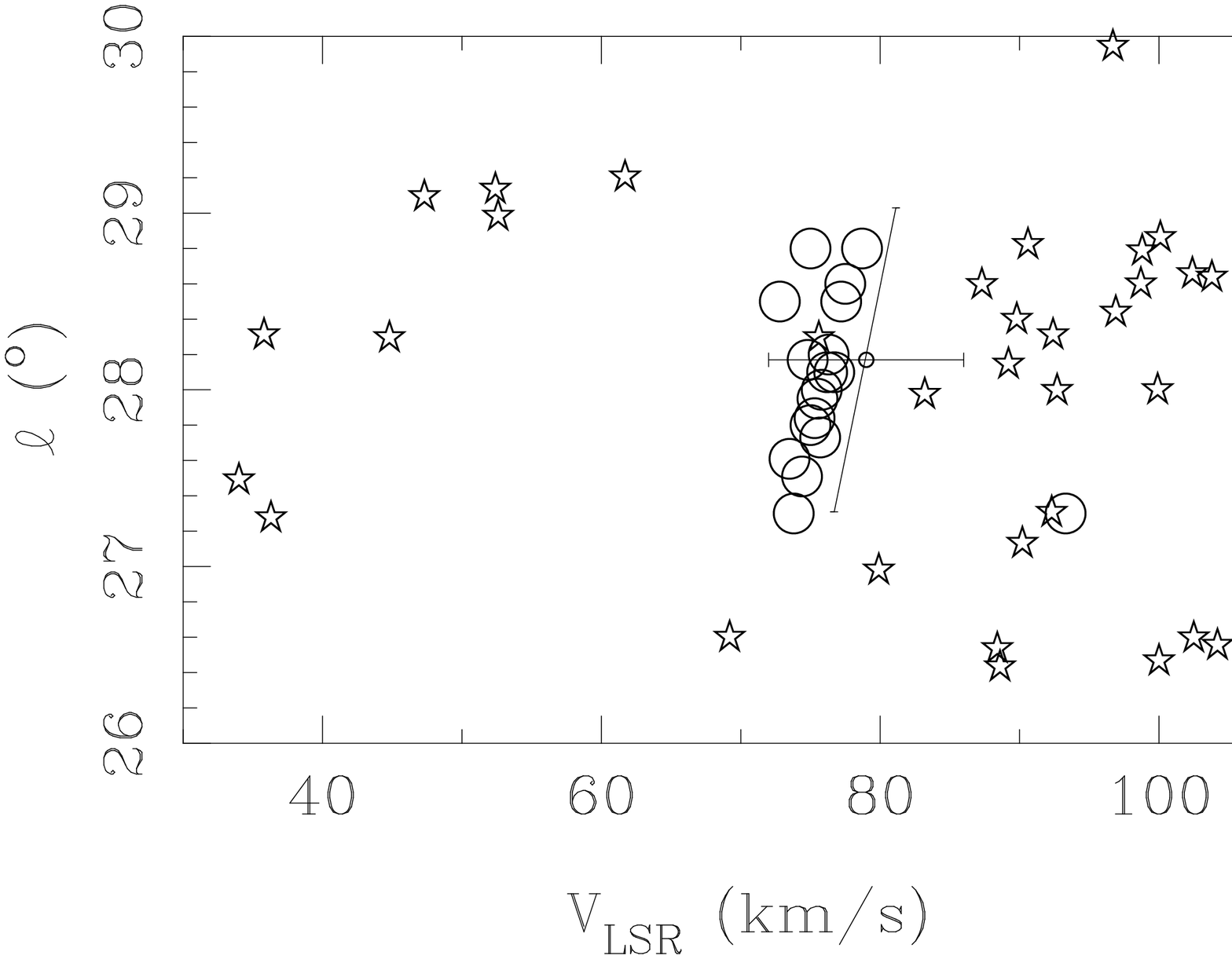}{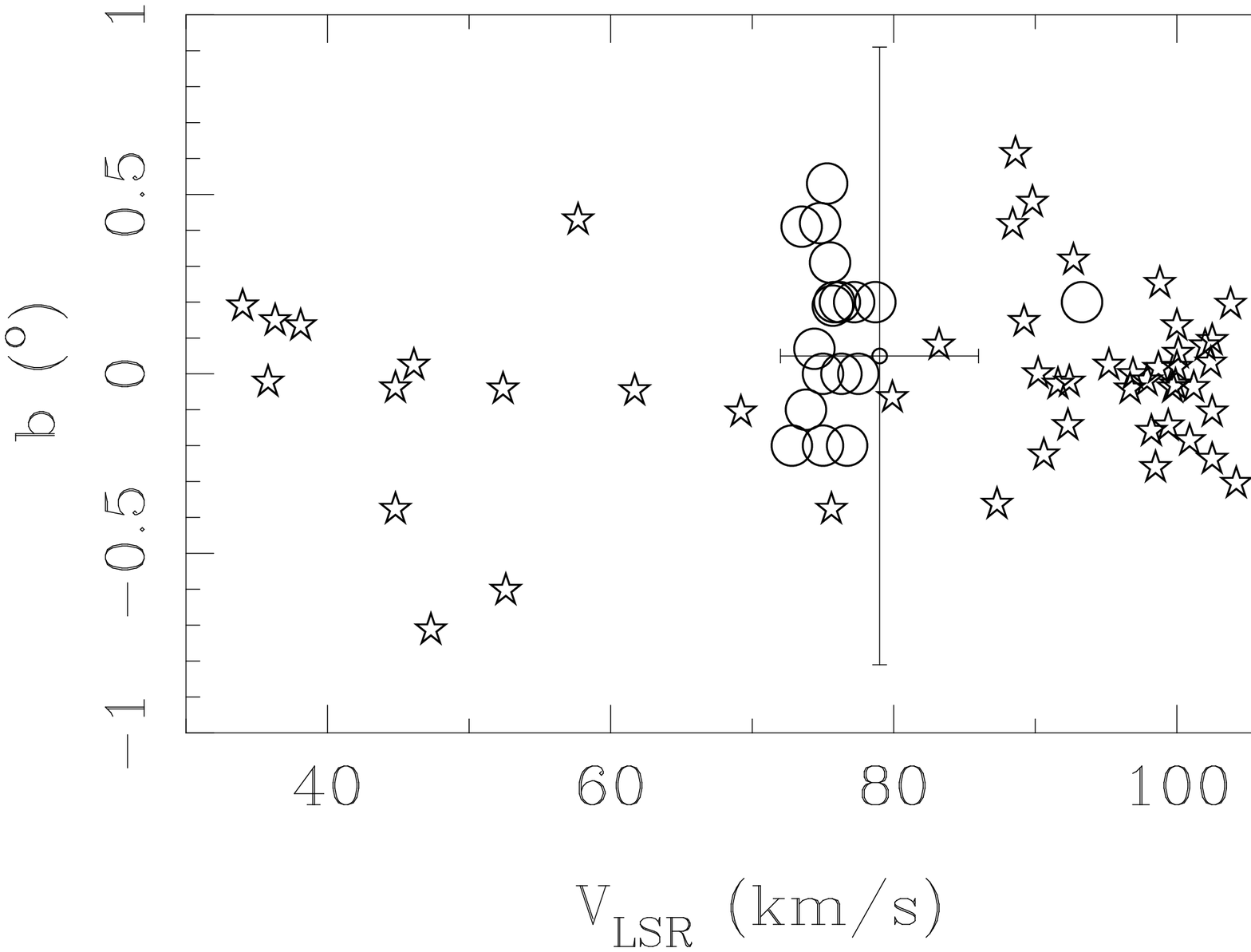}
\epsscale{1.}
\caption{
Velocity-longitude and Velocity-latitude diagrams for the \hi\ 
cloud (large cross), 1720 OH emission (circles) and \hii\ regions
(stars).  There is good correlation between the OH and core of the \hi\ 
cloud, except for a systematic velocity offset of about 2.5 km~s$^{-1}$.
The change in velocity with longitude of both OH and \hi\ is consistent
with the changing sin($l$) projection of Galactic rotation for an 
object at a constant distance from the Galactic Center.
}
\end{figure}

\begin{figure}
\epsscale{1.5}
\plottwo{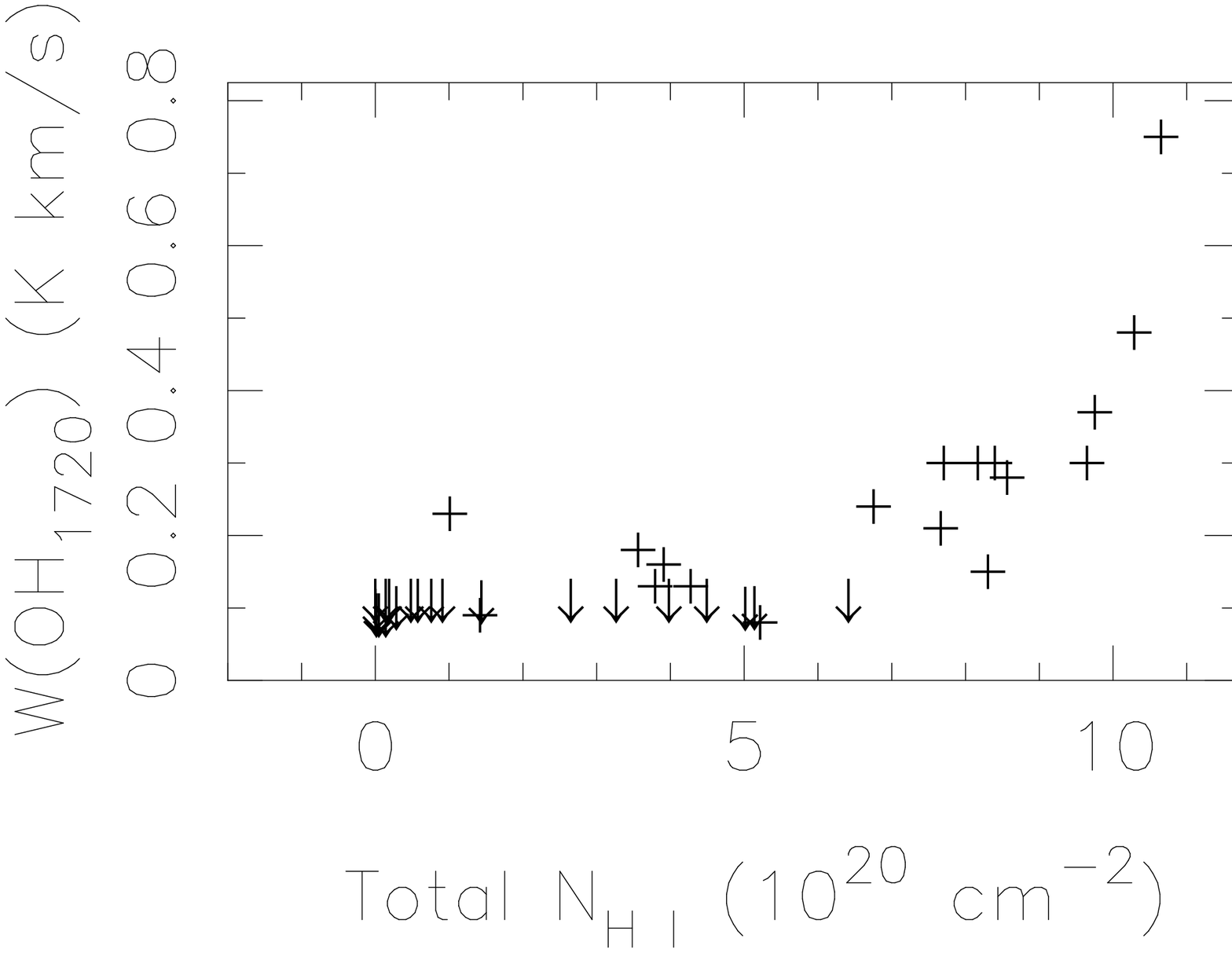}{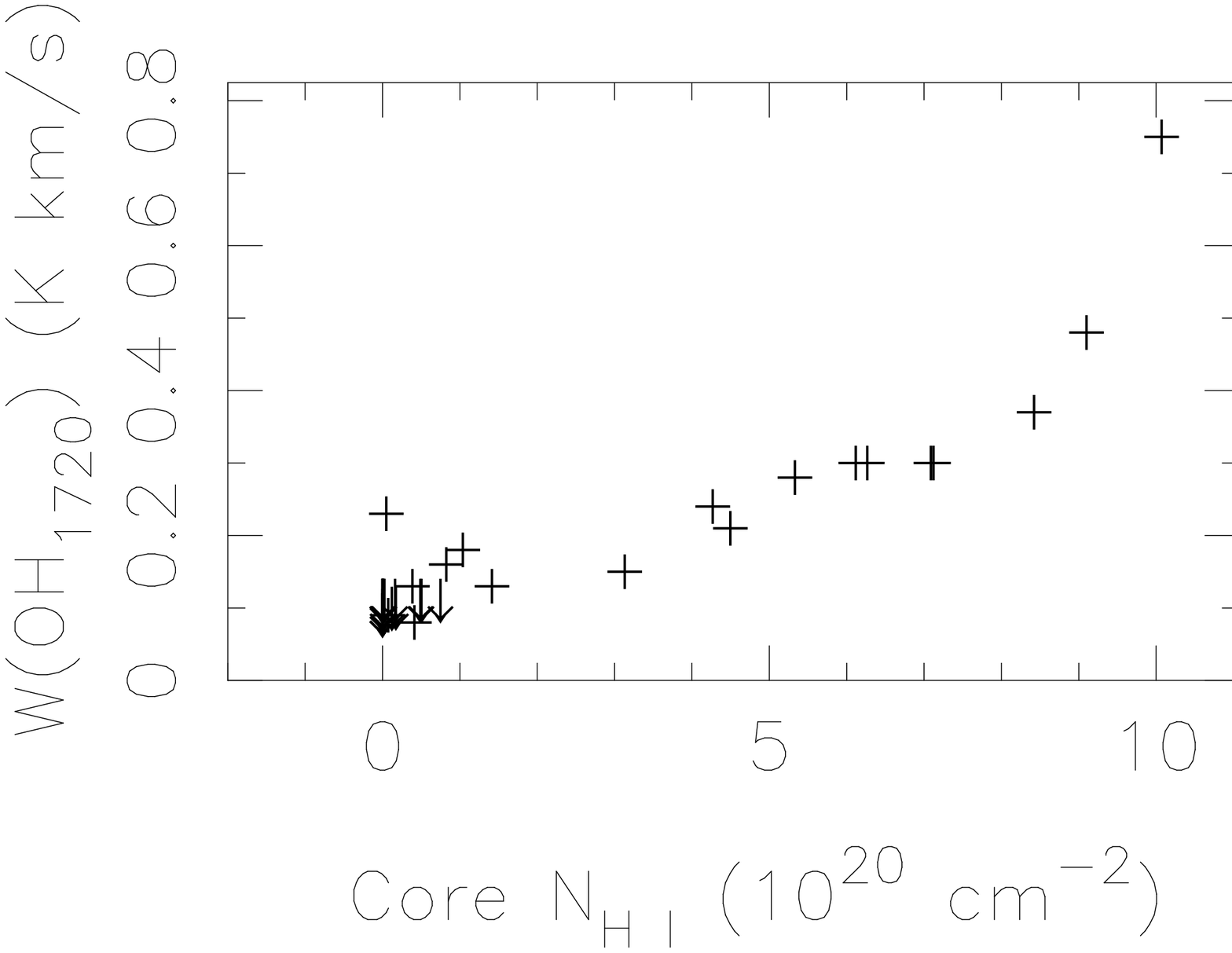}
\epsscale{1.0}
\caption{The integral in the 1720 OH emission line vs. the \hi\ 
column density predicted by Model I.  Upper Panel: Total \hi.  
Lower Panel: \hi\ in the cold core only.  
Anomalous 1720 MHz OH emission appears to be
 correlated with N$_{HI}$ in the cold cloud core
and not with the entire \hi\ cloud.
}
\end{figure}

\begin{figure}
\plotone{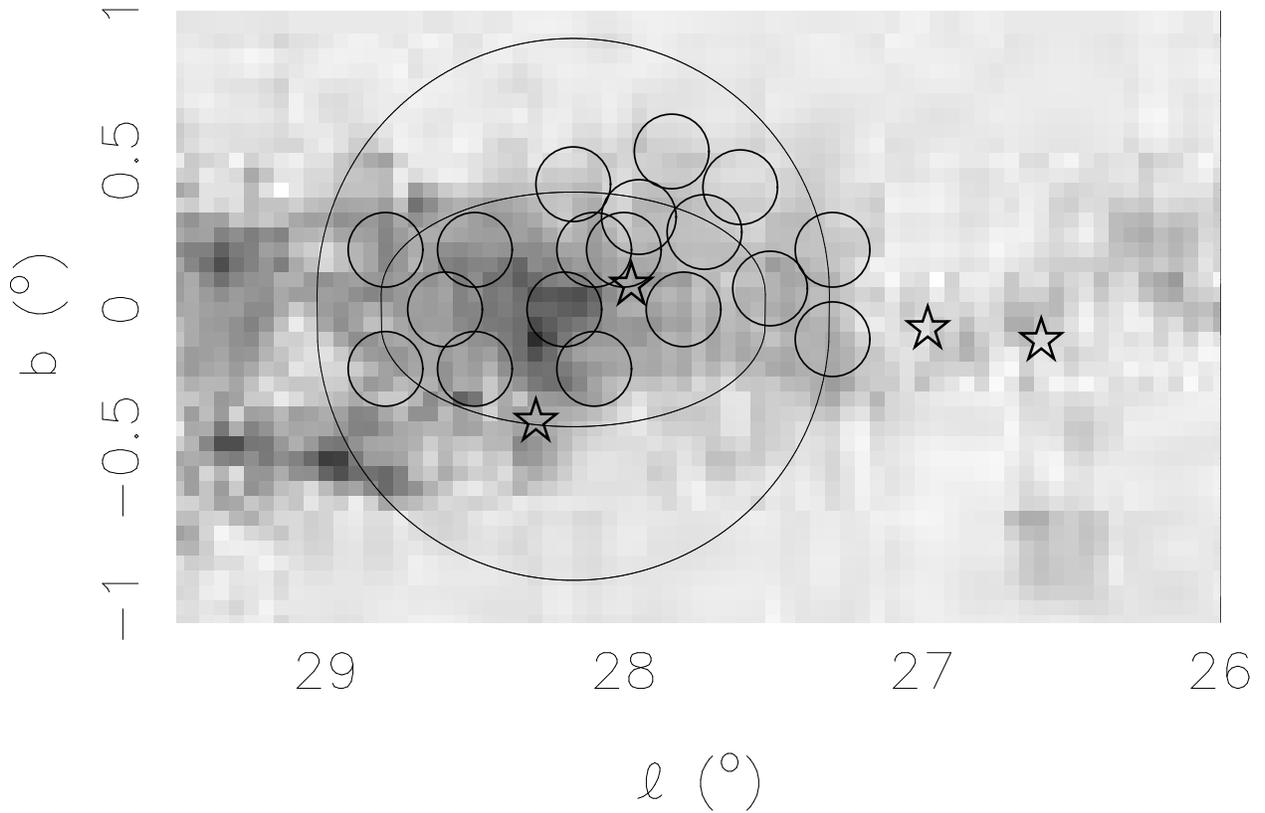}
\caption{The gray scale image is proportional to the $^{12}$CO emission,
the circles mark directions with anomalous 1720 MHz OH emission, and
the stars show \hii\ regions, all within the velocity range of 
the \hi\ cloud, whose outer and core boundaries are also drawn.  The grey scale
runs linearly between 9.23 K~km~s$^{-1}$ (black) and  -0.89 K~km~s$^{-1}$ 
(white).
}
\end{figure}

\begin{figure}
\epsscale{1.5}
\plottwo{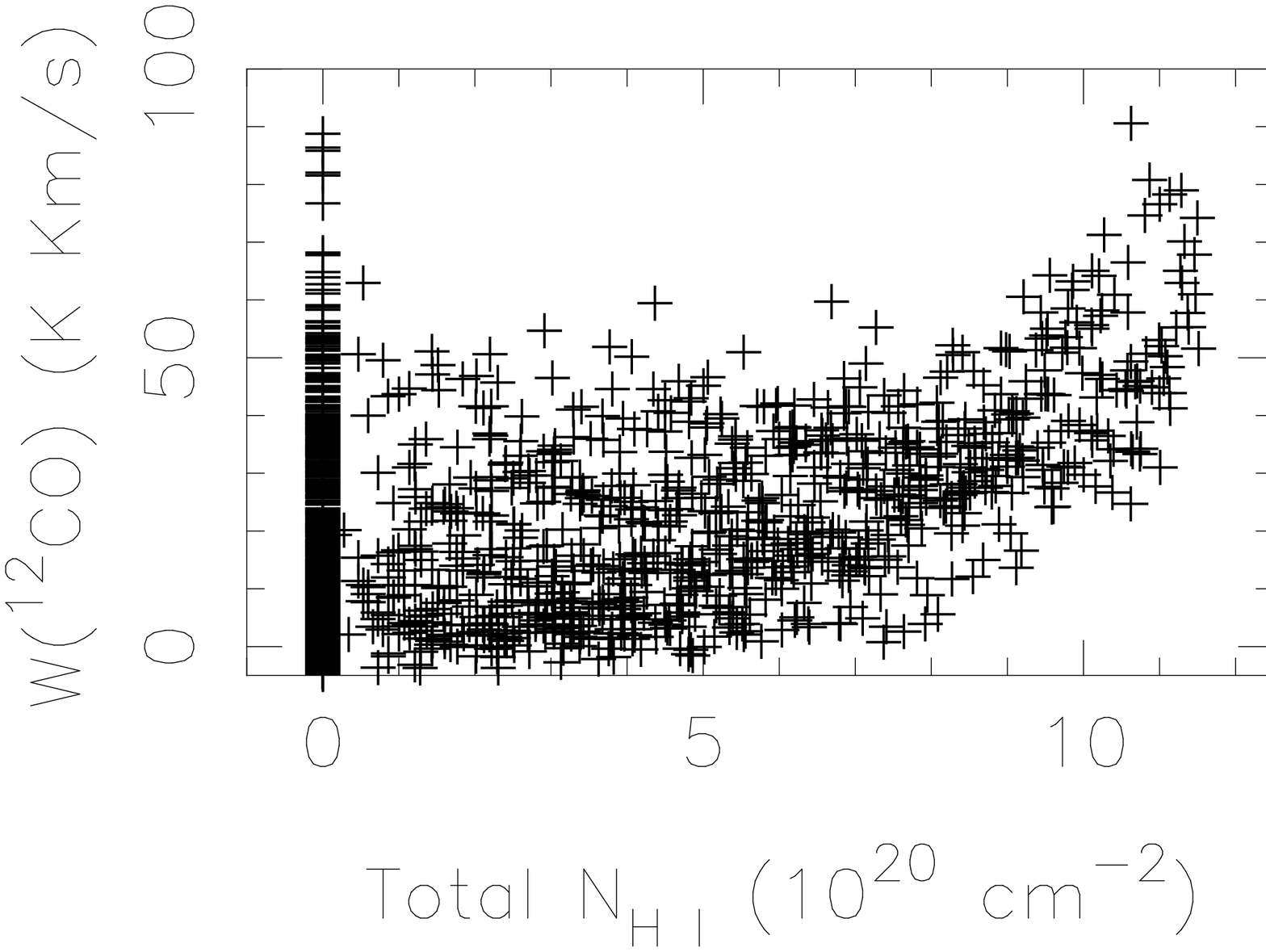}{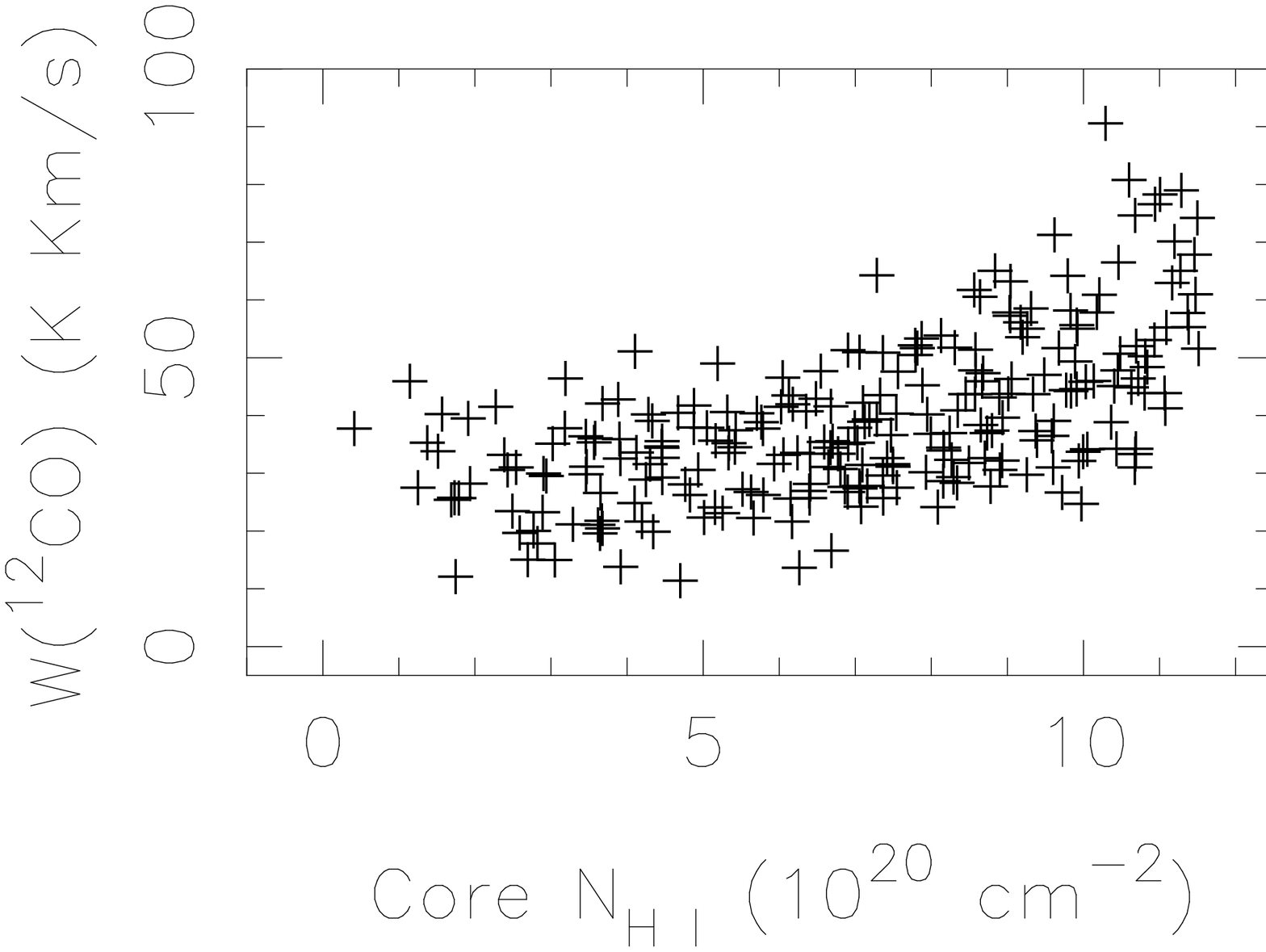}
\epsscale{1.0}
\caption{
Correlation between W($^{12}$CO) and 
${\rm N_{H~I}}$ from Model I.
Upper Panel: All $^{12}$CO in the appropriate velocity range over
longitudes $27\fdg0-29\fdg5$.  Lower Panel: Only the $^{12}$CO towards
the cold \hi\ cloud core.
}
\end{figure}

\begin{figure}
\epsscale{1.0}
\plotone{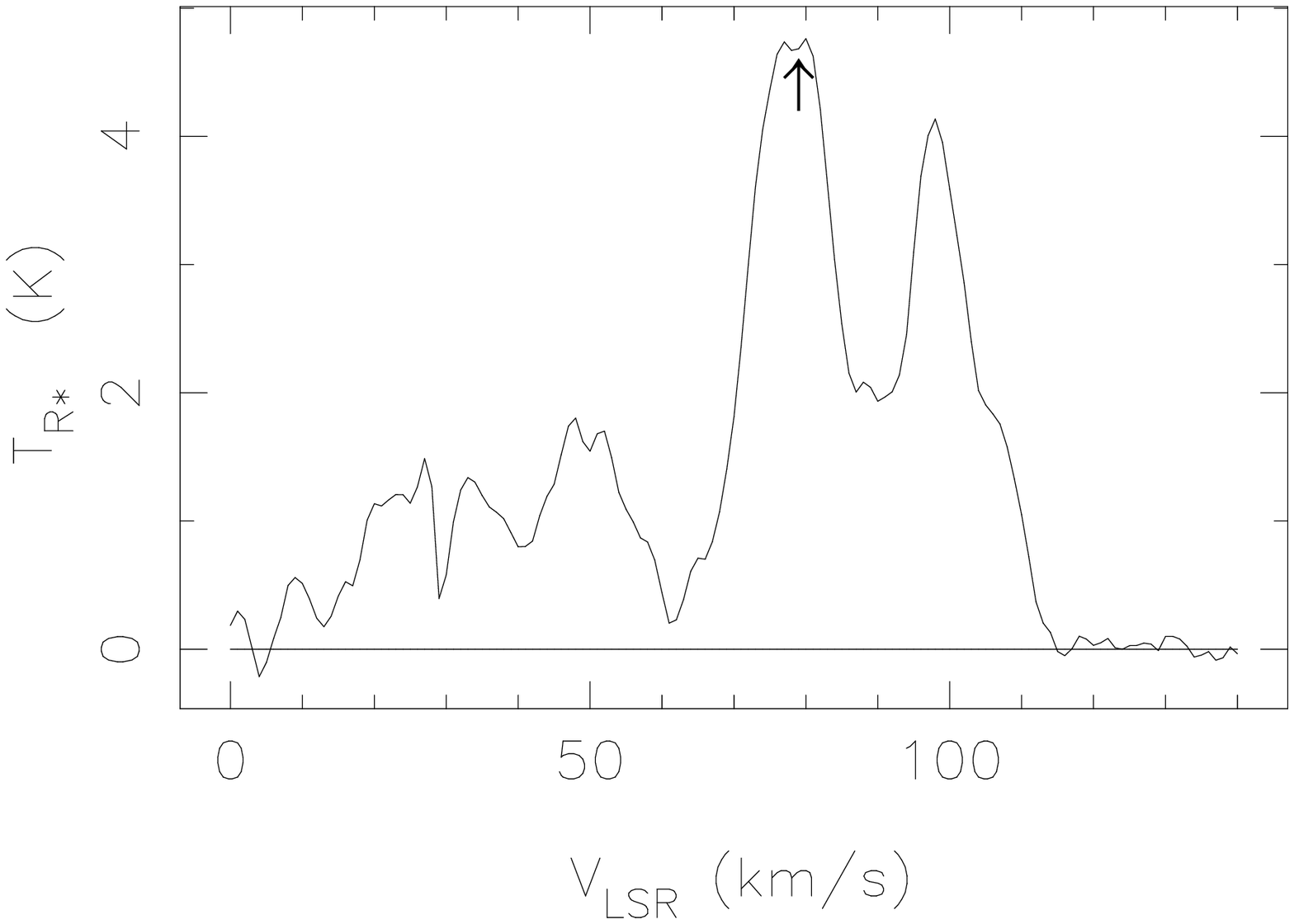}
\caption{Spectrum of $^{12}$CO convolved
to the $21\arcmin$ angular resolution of the \hi\ data
towards the cloud center at $28\fdg17+0\fdg05$ .  The
arrow marks the velocity of the \hi\ cloud.
}
\end{figure}

\end{document}